\documentclass[sigconf]{aamas}
\usepackage{balance}

\usepackage{bm}
\usepackage{algorithm}
\usepackage{algpseudocode}
\usepackage{multirow}
\usepackage{macro}
\usepackage{enumitem}
\usepackage[export]{adjustbox}

\usepackage{fancyhdr}
\renewcommand\footnotetextcopyrightpermission[1]{}

\title[CTRMs for MAPP in Continuous Spaces]{
  CTRMs: Learning to Construct Cooperative Timed Roadmaps\\for Multi-agent Path Planning in Continuous Spaces
}

\author{Keisuke Okumura$^\ast$}
\thanks{$^\ast$Work done during an internship at OMRON SINIC X}
\affiliation{
  \institution{Tokyo Institute of Technology}
  \city{Tokyo}
  \country{Japan}}
\email{okumura.k@coord.c.titech.ac.jp}

\author{Ryo Yonetani, Mai Nishimura}
\affiliation{
  \institution{OMRON SINIC X}
  \city{Tokyo}
  \country{Japan}}
\email{{ryo.yonetani, mai.nishimura}@sinicx.com}

\author{Asako Kanezaki}
\affiliation{
  \institution{Tokyo Institute of Technology}
  \city{Tokyo}
  \country{Japan}}
\email{kanezaki@c.titech.ac.jp}

\begin{abstract}
Multi-agent path planning (MAPP) in continuous spaces is a challenging problem with significant practical importance. One promising approach is to first construct graphs approximating the spaces, called \emph{roadmaps}, and then apply multi-agent pathfinding (MAPF) algorithms to derive a set of conflict-free paths. While conventional studies have utilized roadmap construction methods developed for single-agent planning, it remains largely unexplored how we can construct roadmaps that work effectively for multiple agents. To this end, we propose a novel concept of roadmaps called \emph{cooperative timed roadmaps (CTRMs)}. CTRMs enable each agent to focus on its important locations around potential solution paths in a way that considers the behavior of other agents to avoid inter-agent collisions (\ie, ``cooperative''), while being augmented in the time direction to make it easy to derive a ``timed'' solution path. To construct CTRMs, we developed a machine-learning approach that learns a generative model from a collection of relevant problem instances and plausible solutions and then uses the learned model to sample the vertices of CTRMs for new, previously unseen problem instances. Our empirical evaluation revealed that the use of CTRMs significantly reduced the planning effort with acceptable overheads while maintaining a success rate and solution quality comparable to conventional roadmap construction approaches.$^\dagger$
\end{abstract}
\thanks{$^\dagger$Supplementary material is available at \url{https://omron-sinicx.github.io/ctrm/}}
\renewcommand{\supp}{\n{appendix}}

\begin{document}
\pagestyle{fancy}
\fancyhead{}
\maketitle

\pagestyle{fancy}
\fancyhf{}
\fancyhead[LO,LE]{K. Okumura, R. Yonetani, M. Nishimura, and A. Kanezaki: Learning to Construct CTRMs for Multi-Agent Path Planning}
\cfoot{\vspace{2mm}\thepage\ of \ref{TotPages}}

\section{Introduction}

\emph{Multi-agent path planning (MAPP)} is a fundamental problem in multi-agent systems with attractive applications such as warehouse automation~\cite{wurman2008coordinating} and coordination of self-driving cars~\cite{dresner2008multiagent,khayatian2020survey}. Over the last decade, significant progress has been made on MAPP in \emph{discretized} environments, which is also known as \emph{multi-agent path finding (MAPF)}, aimed at finding a set of conflict-free paths on a graph given \apriori, typically as a grid. Despite its optimization intractability in various criteria~\cite{yu2013structure,ma2016multi,banfi2017intractability}, recent work has shown MAPF can obtain near-optimal solutions in a short amount of time (\eg, less than a minute), even for hundreds of agents~\cite{lam2019branch,li2021eecbs,okumura2021iterative}.

In this work, we tackle another largely unexplored research area: MAPP in \emph{continuous} spaces, which has been recognized as tremendously challenging since the 1980s~\cite{spirakis1984np,hopcroft1984complexity}. On the one hand, allowing agents to move in a continuous space leads to better solutions compared to when their movements are limited in a non-deliberative discretized space such as lattice grids. On the other hand, it is not obvious how to search for these better solutions in the continuous space.

One possible approach to MAPP in continuous spaces consists of two steps: (1)~approximating the continuous spaces by constructing graphs called \emph{roadmaps}, and then (2)~using powerful MAPF algorithms on those roadmaps to derive a solution. While such approaches have been widely used for single-agent planning~\cite{lavalle2006planning}, doing the same for multiple agents is non-trivial. This is primarily due to the necessity of constructing sparse roadmaps, as doing otherwise would make it dramatically more difficult to find a combination of plausible paths due to having to manage a higher number of inter-agent collisions. Nevertheless, there is generally a trade-off between roadmap density and solution quality; roadmaps should be sufficiently dense to ensure a high planning success rate and better solutions. This begs our key question: \emph{``What are the characteristics to consider in order to effectively construct roadmaps for MAPP in continuous spaces?''}

To address this question, we propose a novel concept of graph representations of the space called \emph{cooperative timed roadmaps (CTRMs)}. CTRMs consist of directed acyclic graphs constructed in the following fashion.
(1)~\emph{Agent-specific}: Each CTRM is specialized for individual agents to focus on their important locations. 
(2)~\emph{Cooperative}: Each CTRM is aware of the behaviors of other agents so as to make it easier for the subsequent MAPF algorithms to find collision-free paths.
(3)~\emph{Timed}: Each vertex in CTRMs is augmented in the time direction to represent not only ``where,'' as commonly done in conventional roadmaps, but also ``when,'' because solutions of MAPP are a set of ``timed'' paths. By considering these properties collectively, CTRMs aim to provide a small search space that still contains plausible solutions for MAPF algorithms.

{
  \begin{figure*}
    \centering
    \includegraphics[width=1\linewidth]{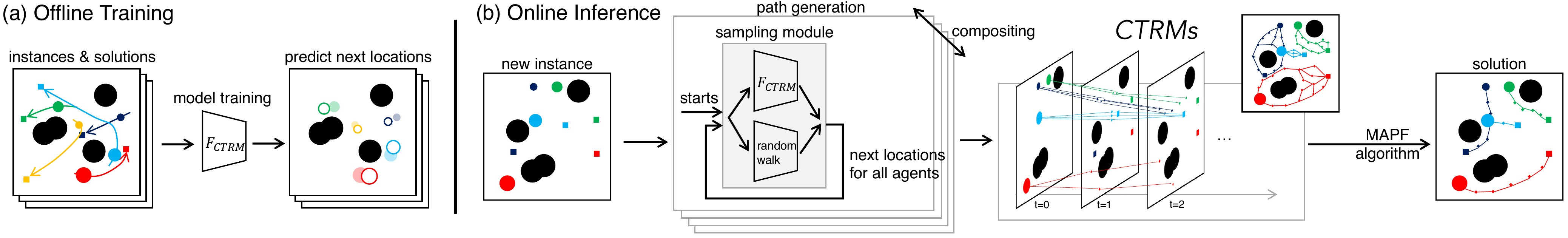}
    \caption{
    Learning to construct CTRMs.
      {\normalfont
      (a) From MAPP demonstrations, we learn a conditional variational autoencoder (CVAE) that predicts how agents move to their goals while avoiding collisions with others. (b) For a new problem instance, we use the learned model as a vertex sampler $F\sub{CTRM}$ to construct CTRMs, on which an MAPF algorithm is invoked to derive solution paths efficiently.
      }
    }
    \label{fig:outline}
  \end{figure*}
}

To construct CTRMs, we develop a machine learning (ML)-based approach (Fig.~\ref{fig:outline}). Suppose we are given a collection of MAPP demonstrations (a pair of problem instances and plausible solution paths) obtained by intensive offline computation with conventional roadmaps, \eg, those constructed with uniform random sampling. The proposed ML model learns from how agents behave cooperatively at each unit of time (\ie, timestep), which is implicitly included in the demonstrations, to predict how they will move in the next timesteps (Fig.~\ref{fig:outline}a). For a new, previously unseen problem instance, the learned model can then be used to sample a small set of agent-specific vertices (\ie, space-time pairs) for generating multiple solution path candidates, and eventually, for constructing CTRMs by compositing those candidates. As a result, the CTRMs can serve a smaller but promising search space and enable the MAPF algorithms to derive a solution more efficiently than when using conventional roadmaps (Fig.~\ref{fig:outline}b). Technically, we extend a class of generative models called the conditional variational autoencoder (CVAE)~\cite{sohn2015learning}, a popular choice for ML sampling-based motion planning~\cite{ichter2018learning,kumar2019lego}, to learn a conditional probability distribution of the vertices of CTRMs for each agent, given the observations of multiple agents. Our model can work with an arbitrary number of agents and even within a heterogeneous setting where agents are diverse in their spatial size and kinematic constraints.

We extensively evaluate our approach on a variety of MAPP problems with several different setups in terms of the number of agents~(21--40), the presence of obstacles, and the heterogeneity in agent sizes and motion speeds. Our results consistently demonstrate that, compared to standard roadmap construction strategies, planning by learning to construct CTRMs is several orders of magnitude more efficient in the planning effort (\eg, assessed by search node expansions or runtime), while maintaining a comparable planning success rate and solution quality, with acceptable overheads. Our main contributions are as follows. 
(1)~We present cooperative timed roadmaps (CTRMs), a novel concept of graph representations tailored to multi-agent path planning in continuous spaces. 
(2)~We develop an ML-based approach to construct CTRMs and demonstrate its effectiveness in solving a variety of MAPP problems.

\section{Related Work}
\label{sec:relatedwork}

Our study is broadly categorized into multi-agent path planning (MAPP), which has evolved in multiple directions and also shares some techniques with single-agent planning, as described below.

\textit{Multi-agent path finding (MAPF)} is a problem of MAPP in discrete spaces, with the objective of finding a set of timed paths on graphs while preventing collisions with other agents~\cite{stern2019def}. MAPF has been studied extensively over the last decade and has produced numerous powerful algorithms (\eg,~\cite{sharon2015conflict,vcap2015prioritized,lam2019branch,okumura2019priority}). More recently, some works have attempted to leverage machine-learning techniques for solving MAPF. These techniques learn from planning demonstrations collected offline to directly predict the next actions of agents given the current observations by means of reinforcement learning~\cite{sartoretti2019primal,damani2021primal,ma2021distributed} or using graph neural networks~\cite{li2020graph,li2021message}. Despite such progress, it remains challenging to determine how these techniques should be applied to MAPP in continuous spaces due to the inherent limitation that assumes the search space to be given \apriori (typically as a grid map).

Path planning in continuous spaces has long been studied for the single-agent case, typically in the form of sampling-based motion planning (SBMP)~\cite{elbanhawi2014sampling} that iteratively samples state points from the space to seek a solution. Popular approaches include probabilistic roadmaps (PRM)~\cite{kavraki1996probabilistic}, rapidly-exploring random trees (RRT)~\cite{lavalle1998rapidly}, or their asymptotically optimal versions (PRM$^\ast$, RRT$^\ast$)~\cite{karaman2011sampling}. Most of the works on MAPP in continuous spaces have been built on these techniques. Roadmaps (or simply pre-defined grid maps) are either constructed or given \apriori to be shared across agents~\cite{van2005prioritized,vcap2015prioritized,honig2018trajectory,yakovlev2017any} or individually for each agent~\cite{solis2021representation,gharbi2009roadmap,kumar2012multi}. A pre-defined set of motion primitives has also been used to discretize possible agent actions~\cite{cohen2019optimal}. Other works extend RRT$^\ast$ to deal with joint state-spaces~\cite{cap2013multi} or the tensor-product of roadmaps constructed for each agent~\cite{shome2020drrt,solovey2016finding}. In contrast to these previous attempts, we explore a different research direction that delves into what roadmap representations are effective for solving MAPP in continuous spaces and how they should be constructed.

Our approach learns how to construct effective roadmaps from MAPP demonstrations. Such a learning-based approach is gaining attention as a promising method for biased sampling in SBMP~\cite{ichter2018learning,ichter2020learned,chen2020learning,qureshi2019motion,kumar2019lego} to sample state points from important regions of given problem instances to derive a solution efficiently. In contrast, only a few studies have examined how learning can improve roadmaps for multi-agent cases. The most relevant work is \cite{felipe2021avoidance}, which proposed learning to bias sampling of PRMs so as to effectively avoid collisions with dynamic obstacles. However, this approach can only be applied to MAPP with homogeneous agents having the same size and kinematic constraints, since it generates a single roadmap shared across all agents. Moreover, it biases samples based on learned obstacle layouts (\eg, narrow passages), making it hard to run in obstacle-free or random-obstacle environments. Another related work~\cite{henkel2020optimized} optimizes already constructed roadmaps by means of online learning, which is orthogonal to the above studies that focus on the roadmap construction.
\section{Preliminaries}
\label{sec:pre}

\paragraph{Notations} We consider a problem of path planning for a team of $N$ agents $\mathcal{A}=\{a_1,\ldots,a_N\}$ in the 2D continuous space $\world \subset \mathbb{R}^2$; hereafter, we simply refer to it as MAPP. Each agent $a_i$ has a body modeled by a convex region $\agentregion_i(q)\in\world$ around its position $q\in\world$, which remains fixed regardless of position but may vary from agent to agent. The space $\world$ may contain a set of convex obstacles \{ $\obstacle_1, \ldots, \obstacle_M \} \subset \world$, where $\mathscr{O}=\bigcup_{j \in \{1, \ldots ,M\}} \obstacle_j$. Then, the obstacle space for an agent $a_i$ is represented by $\mathcal{C}_i^{\rm obs} = \mathscr{O} \ominus \agentregion_i(\bm{0})$, where $\ominus$ is the Minkowski difference and $\bm{0}$ is the origin of $\mathbb{R}^2$. The free space for $a_i$ is $\freespace{i} = \world \setminus \mathcal{C}_i^{\rm obs}$. A trajectory for agent $a_i$ is defined by a continuous mapping $\traj_i:[0, T]\mapsto \world$, where $T\in\mathbb{R}_{>0}$. Each agent $a_i$ has its kinematic constraints $\motion_i$, \eg, each agent has a maximum velocity and moves based on a constant acceleration model.

\paragraph{Problem Description} A \emph{problem instance} of MAPP is defined by a tuple $\mathcal{I}=(\mathcal{A}, \world, \mathscr{O}, \mathscr{R}, \mathscr{K}, \mathcal{S}, \mathcal{G})$, where $\mathscr{R}=\{\agentregion_1,\ldots,\agentregion_N\}$ and $\mathscr{K}=\{\motion_1,\ldots,\motion_N\}$. $\mathcal{S}=\{s_1, \ldots, s_N \mid s_i\in\world\}$ and $\mathcal{G}=\{g_1, \ldots, g_N \mid g_i\in\world\}$ are a set of initial positions and that of goal positions, respectively. A \emph{solution} for the problem instance is a set of $N$ trajectories $\trajs=\{\traj_1,\ldots,\traj_N\}$ that satisfy the following four conditions:
\begin{description}
\item[(Endpoint)] $\traj_i(0) = s_i \land \traj_i(T) = g_i$
\item[(Obstacle)] $\traj_i(\tau) \in \freespace{i}, 0 \leq \tau \leq T$
\item[(Inter-agent)] $\agentregion_i(\traj_i(\tau)) \cap \agentregion_j(\traj_j(\tau)) = \emptyset, i \neq j, 0 \leq \tau \leq T$
\item[(Kinematic-aware)] $\traj_i$ satisfies constraints of $\motion_i$
\end{description}
\label{def:labeled}
Under these conditions, the quality of solution $\trajs$ is measured by \emph{sum-of-costs}: $\sum_i T_i\;\mathrm{where}\; \forall i\in\{1,\ldots,N\},T_i\leq \tau \leq T:\traj_i(\tau)\in g_i$.

\paragraph{Local Planner}
For simplicity, we assume that each agent $a_i$ has a \emph{local planner} $\mu_i: \freespace{i} \times \freespace{i} \times [0, 1] \mapsto \freespace{i} \cup \{ \bot \}$.
This continuous mapping interpolates $a_i$'s locations between distinct space-time pairs under the constraints of $\motion_i$, \eg, $\mu_i(p, q, \epsilon) \defeq (1-\epsilon) p + \epsilon q$ or Dubins paths~\cite{dubins1957curves}, while ensuring $\mu_i(p, q, 0) = p$ and $\mu_i(p, q, 1) = q$.
When such mapping does not exist, the local planner returns $\bot$ as ``not found.''
This is a natural assumption used in planning in continuous space either explicitly or implicitly~\cite{lavalle2006planning}.

\paragraph{Timed Roadmaps} To solve MAPP problems, our approach generates a \emph{timed roadmap} $D_i$ for each agent $a_i$, which approximates the original space \freespace{i} with a finite set of vertices augmented in the time direction, similar to \cite{erdmann1987multiple,fraichard1998trajectory}. Each timed roadmap $D_i$ is defined as a directed acyclic graph $D_i = (V_i, E_i)$, where each vertex $v = (p, t) \in V_i$ is a tuple of space $p \in \world$ and discrete time $t \in \mathbb{N}$, representing that an agent $a_i$ is at location $p$ at timestep $t$. An edge $(u, v) \in E_i$ (to be more precise, arc) exists only when $u=(p, t)$ and $v=(p^\prime, t+1)$. The roadmap $D_i$ is regarded as \emph{consistent} with $a_i$ when (1)~$p \in \freespace{i}$ for all $(p, t) \in V_i$ and (2)~the local planner returns a continuous mapping for all $\left((p, t), (p^\prime, t+1)\right) \in E_i$.

\paragraph{Multi-Agent Path Planner} On the set of timed roadmaps that are consistent with respective agents, $\{D_1,\dots,D_N\}$, a \emph{multi-agent path planner} is invoked to find a set of agent paths defined on discrete and synchronized time: $\paths = ( \path{1}, \ldots, \path{N})$, where $\path{i}=\left(\loc{i}{0}, \ldots, \loc{i}{\lceil T\rceil}\right)\in(\freespace{i})^{\lceil T\rceil}$. We can then obtain a trajectory $\traj_i(\tau)$ that meets the solution conditions by applying the local planner that interpolates between consecutive points in the path $\path{i}$. The local planner is also used to check the inter-agent condition. Any typical MAPF algorithm, such as conflict-based search~\cite{sharon2015conflict} or prioritized planning~\cite{erdmann1987multiple,silver2005cooperative,van2005prioritized,vcap2015prioritized}, is potentially applicable for a multi-agent path planner.
\section{Learning Generative Model for Cooperative Timed Roadmaps}
\label{sec:overview}

{
  \begin{figure}
    \centering
    \includegraphics[width=0.95\linewidth]{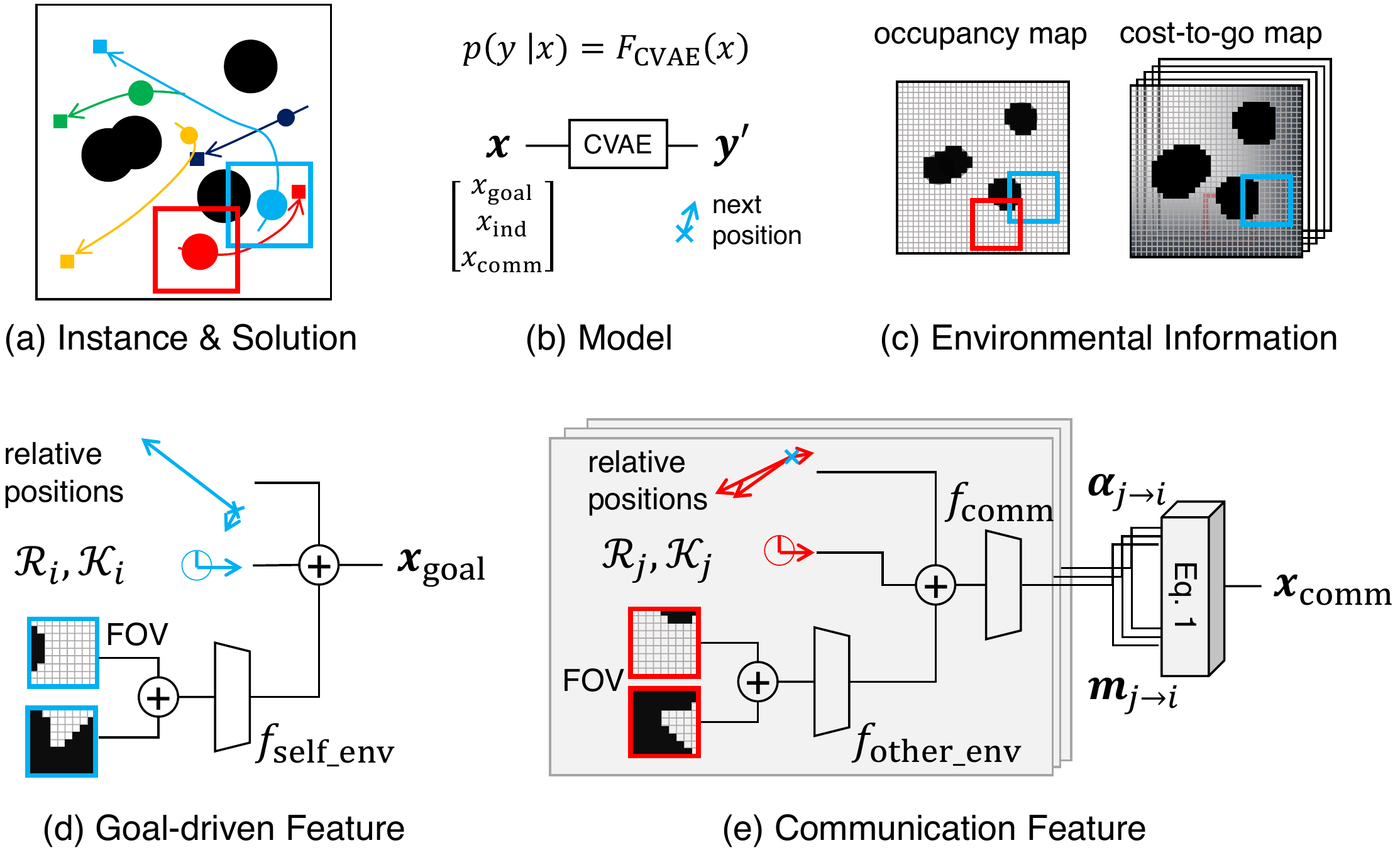}
    \caption{
      The model architecture and its components.
      {\normalfont
      ``$\oplus$'' represents the concatenation of multiple vectors.
      }
    }
    \label{fig:arch}
  \end{figure}
}

Our main technical challenge lies in the construction of timed roadmaps for each agent in a way that effectively reduces the computational effort of multi-agent path planners. We want the roadmaps to provide the planners with a small search space that contains a solution path with plausible quality. To this end, we argue that roadmaps should be ``cooperative'' in that the planners can easily find a conflict-free path by taking into account the presence of other agents during the roadmap construction. This leads to our proposal of CTRMs, which are agent-specific timed roadmaps aware of agent interactions.

\subsection{Model}
In this work, we cast roadmap construction as a machine learning problem. Suppose we are given a collection of MAPP demonstrations consisting of problem instances and their solutions by offline intensive computation of MAPF algorithms on sufficiently dense roadmaps constructed with uniform random or grid sampling (Fig.~\ref{fig:arch}a). Using MAPP demonstrations as training data, we learn a generative model that approximates a probability distribution for vertices constituting a solution path for each agent. The learned model can then be used as a ``vertex sampler'' to construct CTRMs that are more effective for solving a new, previously unseen problem instance efficiently.

As a generative model, we extend the conditional variational auto-encoder (CVAE)~\cite{sohn2015learning}, which is known to be effective for sequence modeling and prediction~\cite{ivanovic2021multimodal,salzmann2020trajectron++}. Our CVAE model takes as input (\ie, condition) a feature vector $\vct{x}$ extracted from the observations of agent $a_i$ and its neighbors at current timestep $t$ in a solution \paths to provide a conditional probability distribution $p(\vct{y}\mid\vct{x})$ of the vector $\vct{y}$ that informs how that agent should move in the next timestep $t+1$ (Fig.~\ref{fig:arch}b). Using a pair of positions $\loc{i}{t}$ and $\loc{i}{t+1}$ extracted from solution path $\path{i}$, we aim to learn the distribution of $\vct{y}=\xi(\loc{i}{t+1}-\loc{i}{t})$, where $\xi(v)=[|v|,(v/|v|)^\top]^\top$ is the magnitude and relative direction of $\loc{i}{t+1}$ with respect to $\loc{i}{t}$. 

Our model consists of two neural networks: an \emph{encoder} and a \emph{decoder}. The encoder ${\it Enc}(\vct{x};\theta)$, parameterized by $\theta$, takes $\vct{x}$ as input to produce a conditional probability distribution $p_\theta(\vct{z}\mid\vct{x})$. The variable $\vct{z}$ is referred to as a \emph{latent variable} that represents $\vct{x}$ in a lower-dimensional space. Then, the decoder ${\it Dec}(\vct{x},\vct{z};\phi)$, parameterized by $\phi$, accepts $\vct{x}$ and $\vct{z}$ drawn from $p_\theta(\vct{z}\mid \vct{x})$ to form another conditional distribution $p_\phi(\vct{y}\mid\vct{x},\vct{z})$. The output distribution $p(\vct{y}\mid\vct{x})$ can be obtained by $p(\vct{y}\mid\vct{x})=\sum_{\vct{z}} p_{\theta}(\vct{z}\mid\vct{x})p_{\phi}(\vct{y}\mid\vct{x},\vct{z})$. Doing so is particularly effective when $\vct{x}$ is high-dimensional, as in our case.

Given a collection of input features $\vct{x}_k$ and ground-truth outputs $\vct{y}_k$, \ie, $\{(\vct{x}_k, \vct{y}_k)\}_{k=1}^K$, extracted across agents and timesteps from problem instances in the training data, we can train the encoder and decoder jointly in a supervised learning fashion (see Sec.~A in the \supp for the details). Once learned, the CVAE can be used as a vertex sampler taking $\vct{x}$ as input to yield a particular sample $\vct{y}'$ in accordance with the learned probability distribution $p(\vct{y}\mid\vct{x})$. In the follows sections, we refer to this sampling process with the learned CVAE as a sampler function $\vct{y}'\sim F\sub{CVAE}(\vct{x})$.

\subsection{Features}
\label{sec:features}
Extracting informative features to form $\vct{x}$ is essential for constructing effective CTRMs. An inherent challenge in MAPP is that each agent needs to move toward its goal while avoiding collisions with other agents that are also in motion. In this section, we first present \emph{goal-driven} features that inform agent $a_i$ of the way to its goal (Sec.~\ref{sec:self}). Then, we introduce \emph{communication} features extracted using an attention network that encodes the information of other agents (Sec.~\ref{sec:comm}). Finally, we propose an \emph{indicator feature} that takes into account high-level choices for which direction to move in when obstacles and other agents are present (Sec.~\ref{sec:ind}).

\subsubsection{Goal-driven Features}
\label{sec:self}
The most basic feature that helps agent $a_i$ move toward its goal $g_i$ is the magnitude and relative direction of $g_i$ with respect to the current position $\loc{i}{t}$, \ie, $\xi(g_i - \loc{i}{t})$. We also include a single-step motion history $\xi(\loc{i}{t-1}- \loc{i}{t})$, parameters of body region $\agentregion_i$, and kinematic constraints $\motion_i$, as they affect how agent $a_i$ moves.

Furthermore, we encode environmental information into the field of view (FOV) of agent $a_i$ around its position $\loc{i}{t}$ to take into account nearby obstacles. We first discretize the original world $\world\subset \mathbb{R}^2$ by the $L^2$ grid, where $L\in\mathbb{N}_{>0}$. We then create the FOV with size $l^2$, where $l \in \mathbb{N}_{> 0}$, centered around $\loc{i}{t}$. Within the FOV, we extract two binary maps (illustrated in Fig.~\ref{fig:arch}c: \emph{local occupancy map}) indicating if each grid cell is occupied by obstacles and a \emph{cost-to-go-map} that shows if each cell is closer to the goal compared to the current position, pre-computed by means of a breadth-first search on the grid. We feed these two maps to a neural network $\nn{self\_env}$ to transform them into a compact vector. As shown in Fig.~\ref{fig:arch}d, we then concatenate these features to form a goal-driven feature vector $\vct{x}\sub{goal}$.

\subsubsection{Communication Features}
\label{sec:comm}
To learn the cooperative behaviors between agents observed in ground-truth paths, it is critical to extract features about other agents $a_j$, such as where they are in the current and previous timesteps, \ie, $\xi(\loc{j}{t}-\loc{i}{t}), \xi(\loc{j}{t-1}-\loc{i}{t})$, where they will be moving, \ie, their goal position $\xi(g_j-\loc{i}{t})$, and how their trajectory could be affected by agent body $\agentregion_j$ and kinematic constraints $\motion_j$. We also take into account the occupancy and cost-to-go maps around the position of $a_j$ by feeding them into a neural network $\nn{other\_env}$ to be represented by a compact vector. Let $\vct{x}_{j\rightarrow i}$ be a feature vector concatenating all this information about $a_j$ with respect to $a_i$, as shown in Fig.~\ref{fig:arch}e.

Now suppose that we are given a collection of features $\{\vct{x}_{j\rightarrow i}\mid j \in \neigh{i}\}$ for all the agents nearby target $a_i$, where $\neigh{i}$ is a set of indices for the predefined number of neighboring agents of $a_i$. The question then is how to aggregate them as a part of feature vector $\vct{x}$. Obviously, just concatenating them all is not scalable and would not be able to deal with problem instances affecting variable numbers of agents. Instead, we leverage the recent progress in multi-agent interaction modeling that deals with agent communications using an attention network~\cite{hoshen2017vain}. Specifically, we feed $\vct{x}_{j\rightarrow i}$ to a neural network $\nn{comm}$ that outputs two variables: attention vector $\vct{\alpha}_{j\rightarrow i}$ and message vector $\vct{m}_{j\rightarrow i}$. Message vectors are then aggregated while weighted using attention vectors to provide a communication feature vector $\vct{x}\sub{comm}$, as follows:

\begin{equation}
  \begin{split}
  \vct{x}\sub{comm} &= \sum_{j \in \neigh{i}} \vct{m}_{j\rightarrow i}\cdot w_{j\rightarrow i},
  \\
  w_{j \rightarrow i} &= \frac{\exp \left[-\|\vct{\alpha}_{j\rightarrow i} - \vct{\alpha}_{i\rightarrow i}\|^2 \right]}{\sum_{k\in\neigh{i}}\exp \left[-\|\vct{\alpha}_{k\rightarrow i} - \vct{\alpha}_{i\rightarrow i}\|^2 \right]}
  \end{split}
\label{eq:vain}
\end{equation}
where $w_{j\rightarrow i}$ is a scalar weight for the message vector $\vct{m}_{j\rightarrow i}$, which is defined by the L2 distance between two attention vectors $\vct{\alpha}_{j\rightarrow i}$ and $\vct{\alpha}_{i\rightarrow i}$ normalized across $j\in\neigh{i}$ using the softmax function. With Eq.~(\ref{eq:vain}), agent $a_i$ can consider message vectors only from selected agents who are close to $a_i$ in terms of attention vectors.

{
  \begin{figure}
    \centering
    \includegraphics[width=0.90\linewidth]{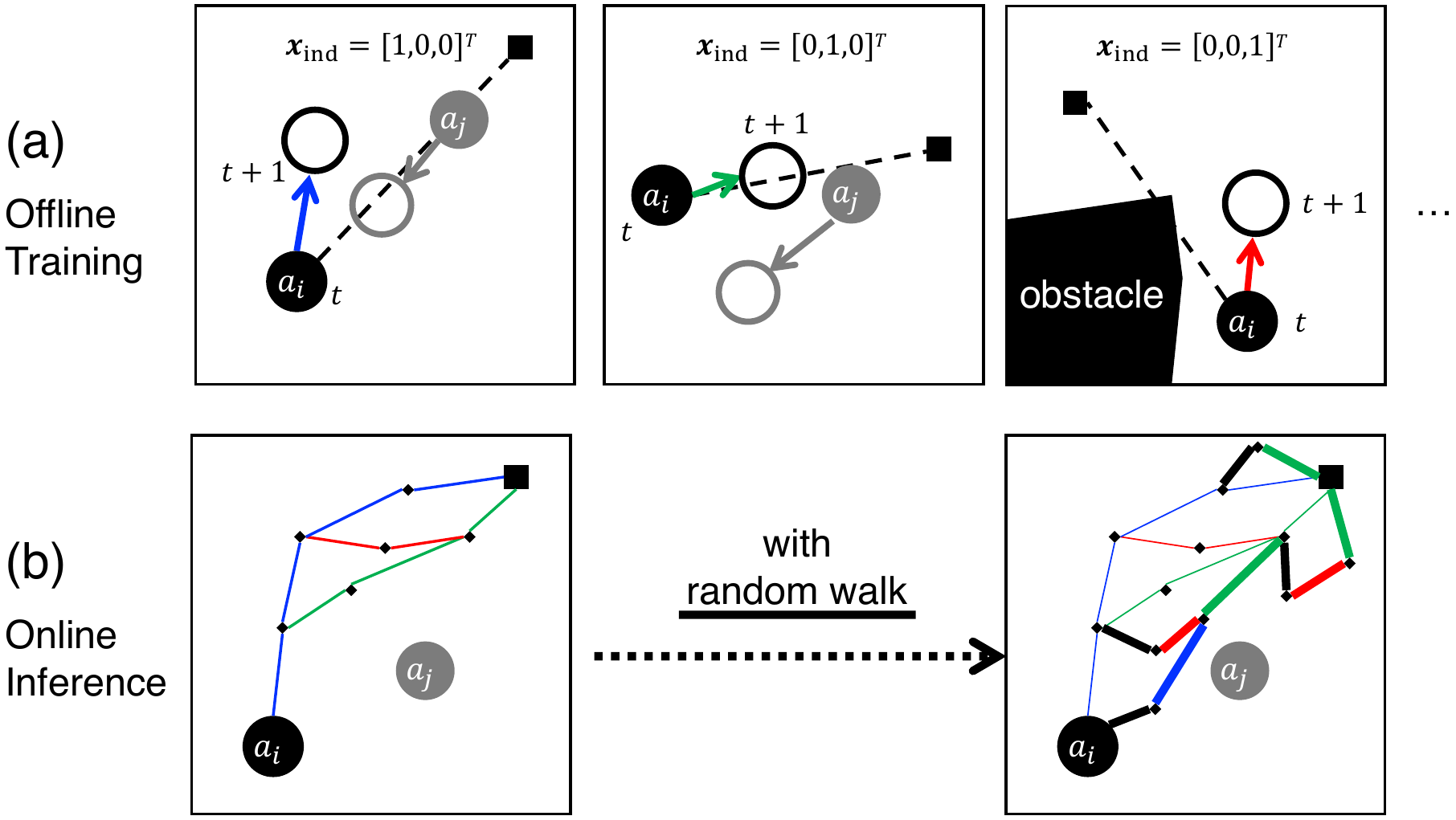}
    \caption{
      Illustration of the indicator function. Agents are represented by solid black/gray circles.
    }
    \label{fig:indicator}
  \end{figure}
}

\subsubsection{Indicator feature}
\label{sec:ind}
Figure~\ref{fig:indicator}a depicts typical situations where we want to sample the next motion of target agent $a_i$ in the presence of another agent $a_j$ or obstacles in front of $a_i$. As shown in these examples, agents are often presented with multiple and discrete choices about their moving direction (such as moving left, straight, or right) depending on the layout of the surrounding agents and obstacles. As such, we expect the sampling of next motions to be affected by such high-level choices, which is indeed crucial for constructing effective CTRMs, as we will empirically demonstrate in our experiments.

To this end, we propose augmenting an input feature $\vct{x}$ with a discrete feature called an \emph{indicator feature} $\vct{x}\sub{ind}$ that is learned to indicate promising choices of the next moving directions based on the current observations. Specifically, we define $\vct{x}\sub{ind}$ by the relative direction of $\loc{i}{t+1}$ from $\loc{i}{t}$ with respect to the goal direction $g_i-\loc{i}{t}$, in the one-hot form such as left, straight, or right, respectively indicated by $[1, 0, 0]^\top$, $[0, 1, 0]^\top$, and $[0, 0, 1]^\top$ (concrete implementations are presented in Sec.~\ref{sec:impl}.) While calculating this feature from ground-truth path \path{i} in the training phase, we also learn a neural network \nn{ind} taking the concatenation of $\vct{x}\sub{goal}$ and $\vct{x}\sub{comm}$ as input to predict it. The learned network is then used in the inference phase to provide $\vct{x}\sub{ind}$, where $\loc{i}{t+1}$ is unknown.

\subsubsection{Putting Everything Together}
\label{sec:together}
Finally, we concatenate all features to form $\vct{x}$, \ie, $\vct{x}=[x\sub{goal}^\top, \vct{x}\sub{comm}^\top, x\sub{ind}^\top]^\top$. Note that the neural networks used for the feature extraction, \ie, $\nn{self\_env}$, $\nn{other\_env}$, $\nn{comm}$, and $\nn{ind}$, can be trained end-to-end with the CVAE.

\section{Constructing Cooperative Timed Roadmaps with Learned Model}
\label{sec:roadmap-generation}

\subsection{Algorithm Overview}
\begin{algorithm}[t]
  \caption{\funcname{construct\_CTRMs}}
  \label{algo:roadmap-generation}
  \begin{algorithmic}[1]
  \item[\textbf{Input}:~a problem instance \I]
  \item[\textbf{Output}:~a set of timed roadmaps $\mathcal{D} = \{D_1, \ldots, D_N\}$]
  \item[\textbf{Hyperparameters}: $\ntraj \in \mathbb{N}_{>0}, \tmax \in \mathbb{N}_{>0}$]
    \State $D_i \leftarrow \left(\{ (s_i, 0) \}, \emptyset \right)$: for each $i = 1, \ldots, N$
    \State $\tmakespan \leftarrow 0$
    \For{$j=1, \ldots, \ntraj$}
    \label{algo:traj:iter-start}
    \State initialize table \lochead
    \label{algo:traj:init-table}
    \Comment \locbody{i}{t} is location of $a_i$ at timestep $t$
    \State $\locbody{i}{0} \leftarrow s_i$: for each $i = 1, \ldots, N$
    \label{algo:traj:insert-start}
    \For{$t=1, \ldots, \tmax - 1$}
    \Comment create one set of trajectories
    \label{algo:traj:loop-tmax-start}
    \For{$i=1, \ldots, N$}
    \State $p \leftarrow \funcname{sample\_next\_vertex}\left(\I, t, i, \lochead\right)$
    \label{algo:traj:sample}
    \Comment $p \in \freespace{i}$
    \State $q \leftarrow \funcname{find\_compatible\_vertex}(\I, t, i, p, D_i)$
    \label{algo:traj:merge}
    \If{$q$ is not found}
    \State $\locbody{i}{t} \leftarrow p$; $\funcname{insert}(p, t, D_i)$
    \label{algo:traj:insert}
    \Else
    \State $\locbody{i}{t} \leftarrow q$
    \label{algo:traj:update-by-q}
    \EndIf
    \EndFor
    \If{$\funcname{check\_reachability\_to\_goals}(\I, t, \lochead)$}
    \label{algo:traj:check-goal}
    \State $\tmakespan \leftarrow \max\left\{T + 1, \tmakespan\right\}$; \textbf{break}
    \label{algo:traj:stop-path-gen}
    \EndIf
    \EndFor
    \label{algo:traj:loop-tmax-end}
    \EndFor
    \label{algo:traj:iter-end}
    \State $\funcname{insert}(g_i, t, D_i)$: for each $t = 1, \ldots, \tmakespan $, $i = 1, \ldots, N$
    \label{algo:traj:insert-goal}
    \State \Return $\mathcal{D} = \{D_1, \ldots, D_N\}$
  \end{algorithmic}
\end{algorithm}
In this section, we explain how to construct CTRMs using the learned model as a vertex sampler. Algorithm~\ref{algo:roadmap-generation} is our proposed method which takes a problem instance \I as input to build a set of CTRMs consistent with respective agents, \ie, $\mathcal{D}=\{D_1,\dots,D_N\}$.

We generate a sequence of vertices (\ie, path) with a maximum length $\tmax \in \mathbb{N}_{>0}$ for each agent using \funcname{sample\_next\_vertex} presented in Sec.~B.1 of the \supp, which is repeated $\ntraj \in \mathbb{N}_{>0}$ times to construct CTRMs. Each path being generated is temporally stored in the table \lochead, where $\locbody{i}{t}$ refers to the location of agent $a_i$ sampled for timestep $t$. This table is used by the learned model in \funcname{sample\_next\_vertex} to sample the next vertices for each agent while being aware of the locations of other agents; we use $\locbody{i}{t}$ as a proxy of $\loc{i}{t}$ for the feature extraction described in Sec.~\ref{sec:features}.

During each path generation, we invoke the \funcname{insert} function to add a sampled vertex $(p, t)$ to the current CTRM $D_i = (V_i, E_i)$ and update $E_i$ properly (Line~\ref{algo:traj:insert}). This procedure is done only when we confirm that the sampled vertex is not compatible with those we already have in $V_i$ using \funcname{find\_compatible\_vertex}, as explained in Sec.~\ref{sec:merge}. Further, we keep updating $T\sub{makespan}$ to be the smallest number of timesteps to move all the agents to their goals by checking whether they can reach their goals from their last locations with the \funcname{check\_reachability\_to\_goals} function (Line~\ref{algo:traj:check-goal}). After all the iterations, the algorithm inserts each agent's goal $g_i$ into $D_i$ for time $t=1,\ldots,T\sub{makespan}$ (Line~\ref{algo:traj:insert-goal}) and returns a set of CTRMs $\mathcal{D}$.

\subsection{Combining Sampling from Learned Model with Random Walk}
\label{sec:sampling}

In the sub-routine \texttt{sample\_next\_vertex}, we leverage a learned model $F\sub{CTRM}$ to sample vertices such that CTRMs can efficiently cover a variety of possible solution paths. The complete algorithm description is presented in Sec.~B.1 of the \supp.

The crucial point here is that we replace the learned model with a random walk centered at the current location (\locbody{i}{t-1}) at probability $1 - \pbiased$. This contributes to improving the expressiveness of CTRMs beyond what has been learned from the training data, as illustrated in Fig. \ref{fig:indicator}b. Similar techniques are commonly introduced in learning-based sampling for SBMP~\cite{ichter2018learning,ichter2020learned,chen2020learning}. We set \pbiased to be low in the initial timestep and gradually increase it from there.

{
    \setlength{\tabcolsep}{0pt}
    \newcommand{\figcol}[2]{
    \begin{minipage}[t]{0.185\linewidth}
    \centering
    {\small #2}\\
    \includegraphics[width=0.9\linewidth]{fig/raw/roadmaps/roadmap_#1.pdf}
    \end{minipage}
}

\begin{figure*}
\centering
\scalebox{0.92}{
\begin{tabular}{ccccc}
    \figcol{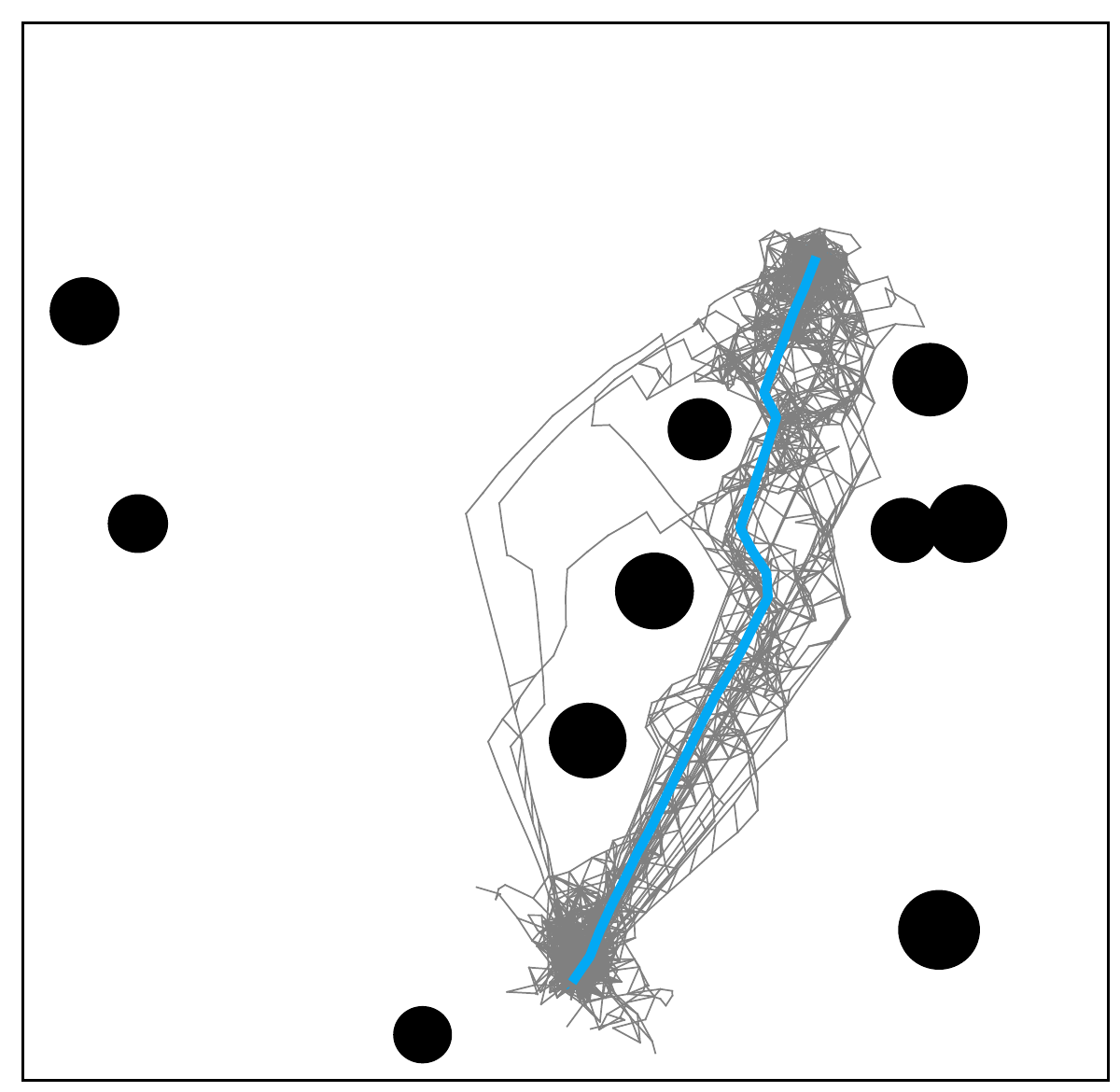}{CTRMs}
    \figcol{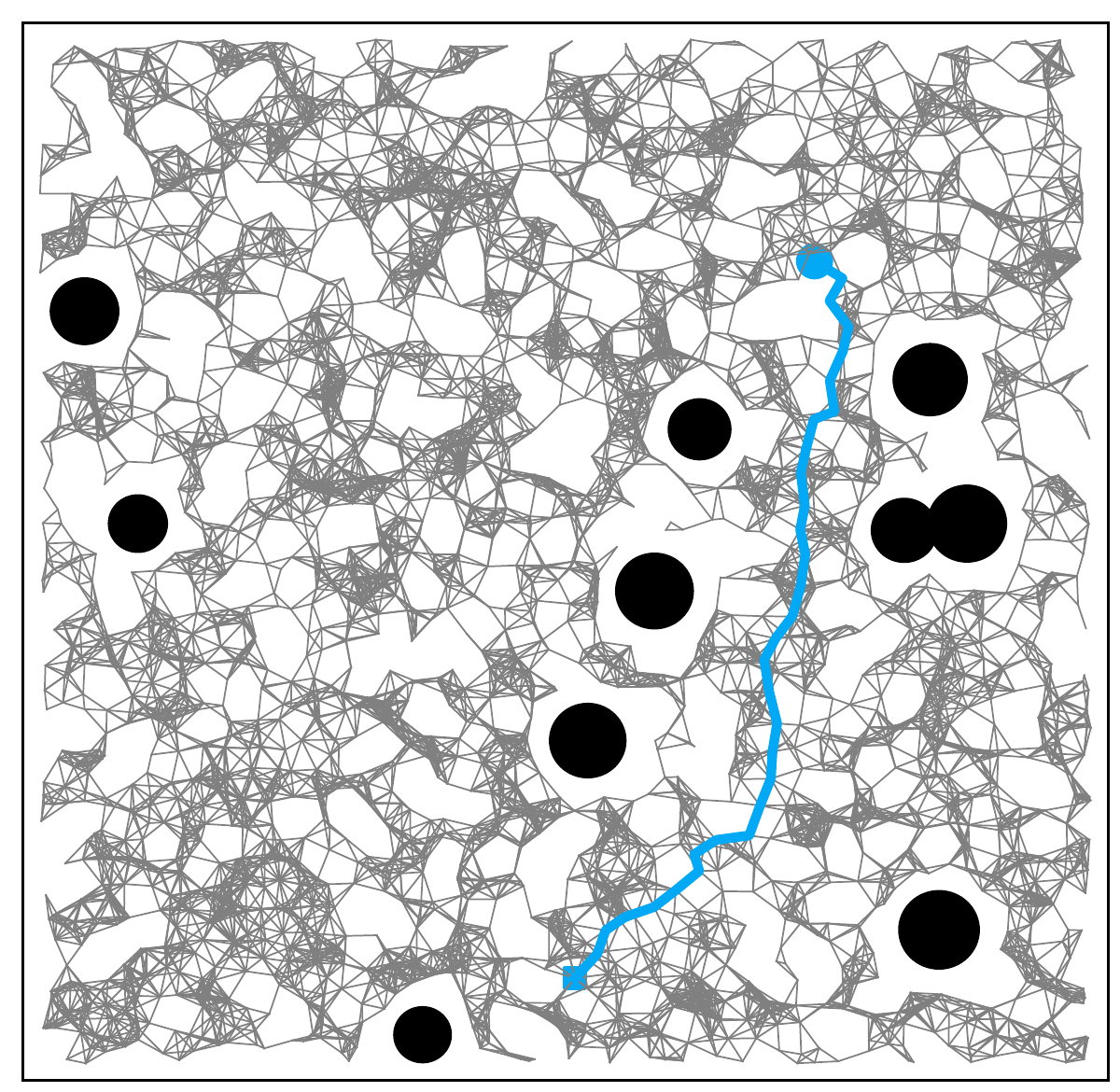}{random}
    \figcol{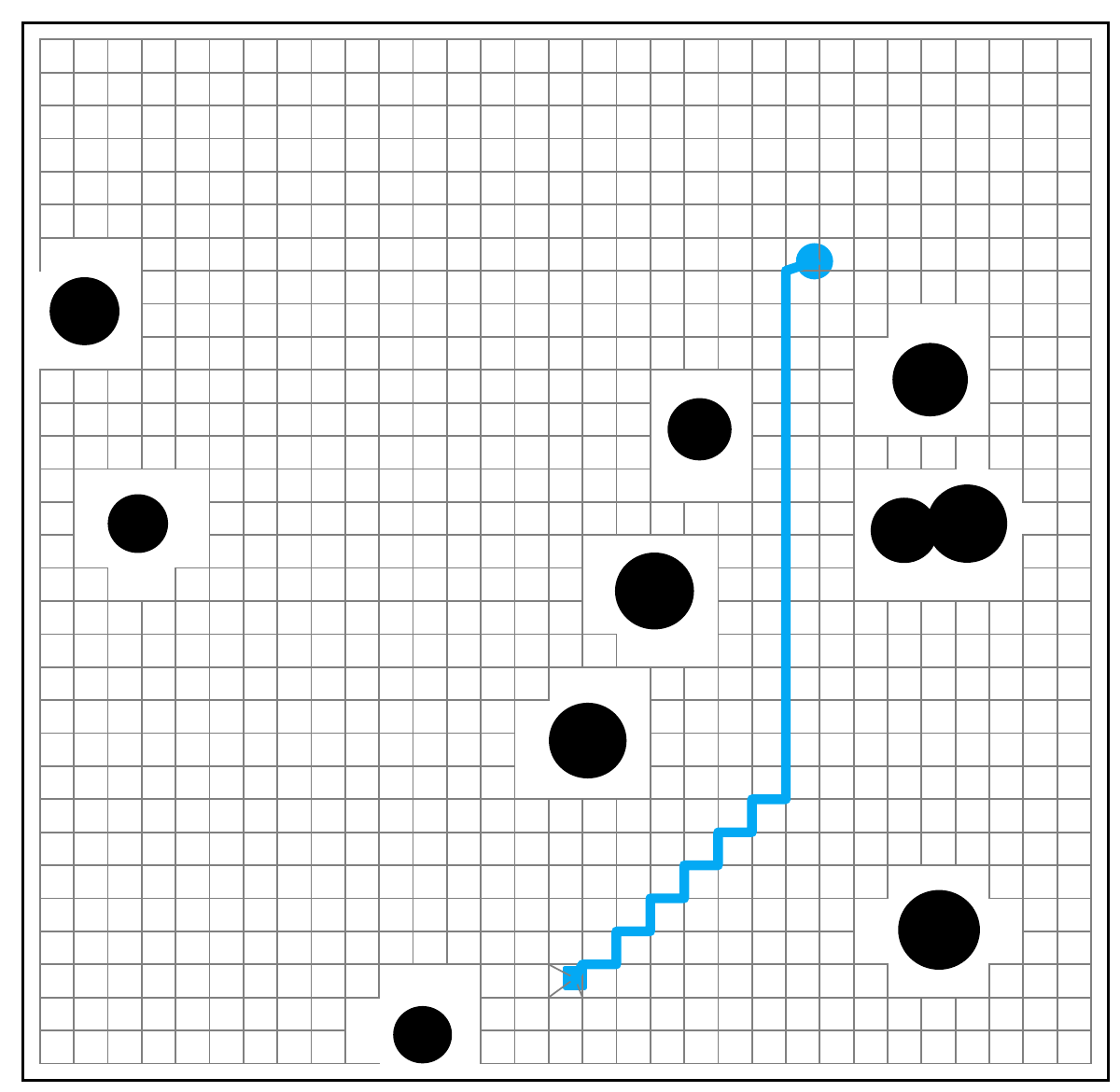}{grid}
    \figcol{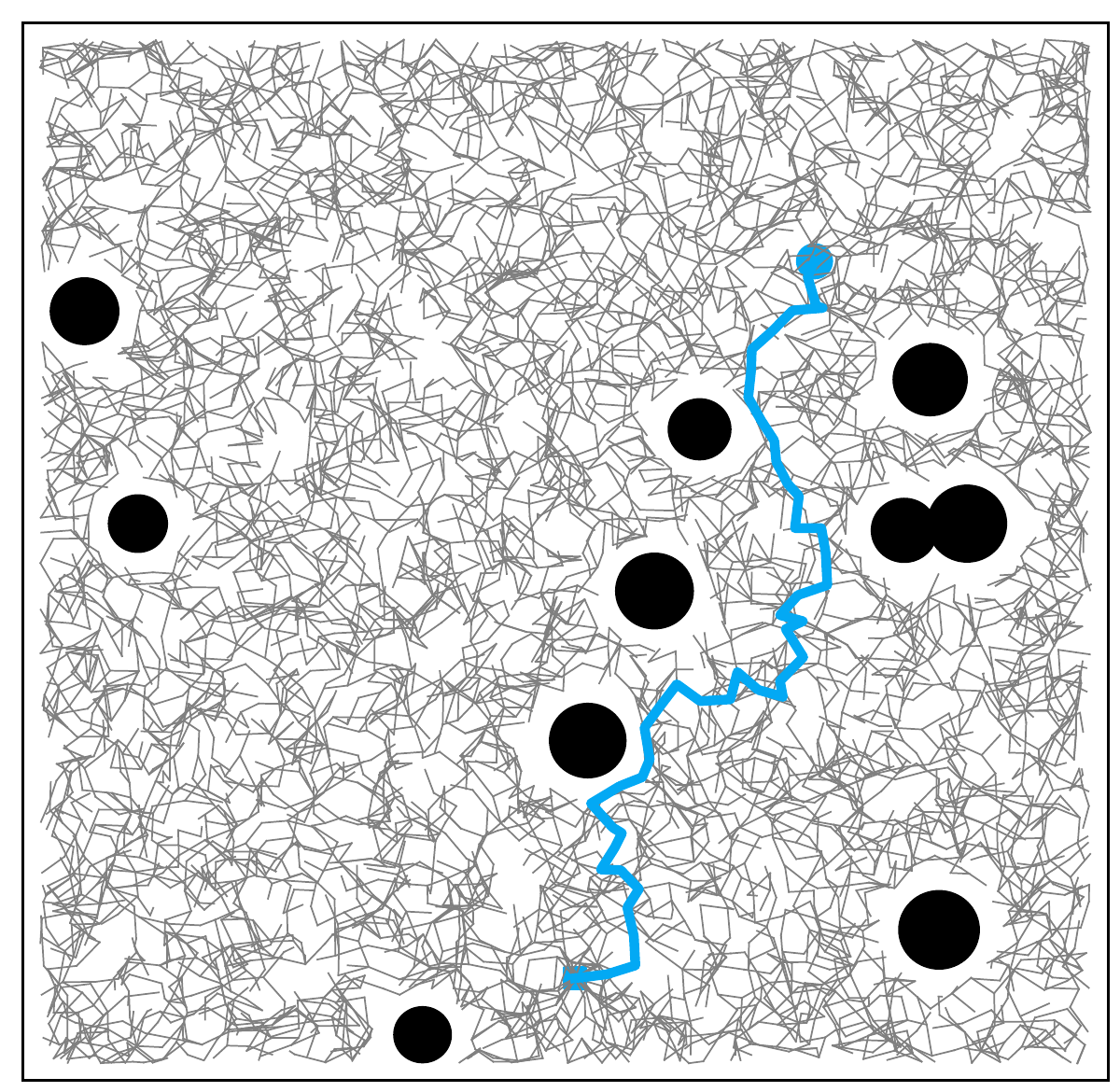}{SPARS}
    \figcol{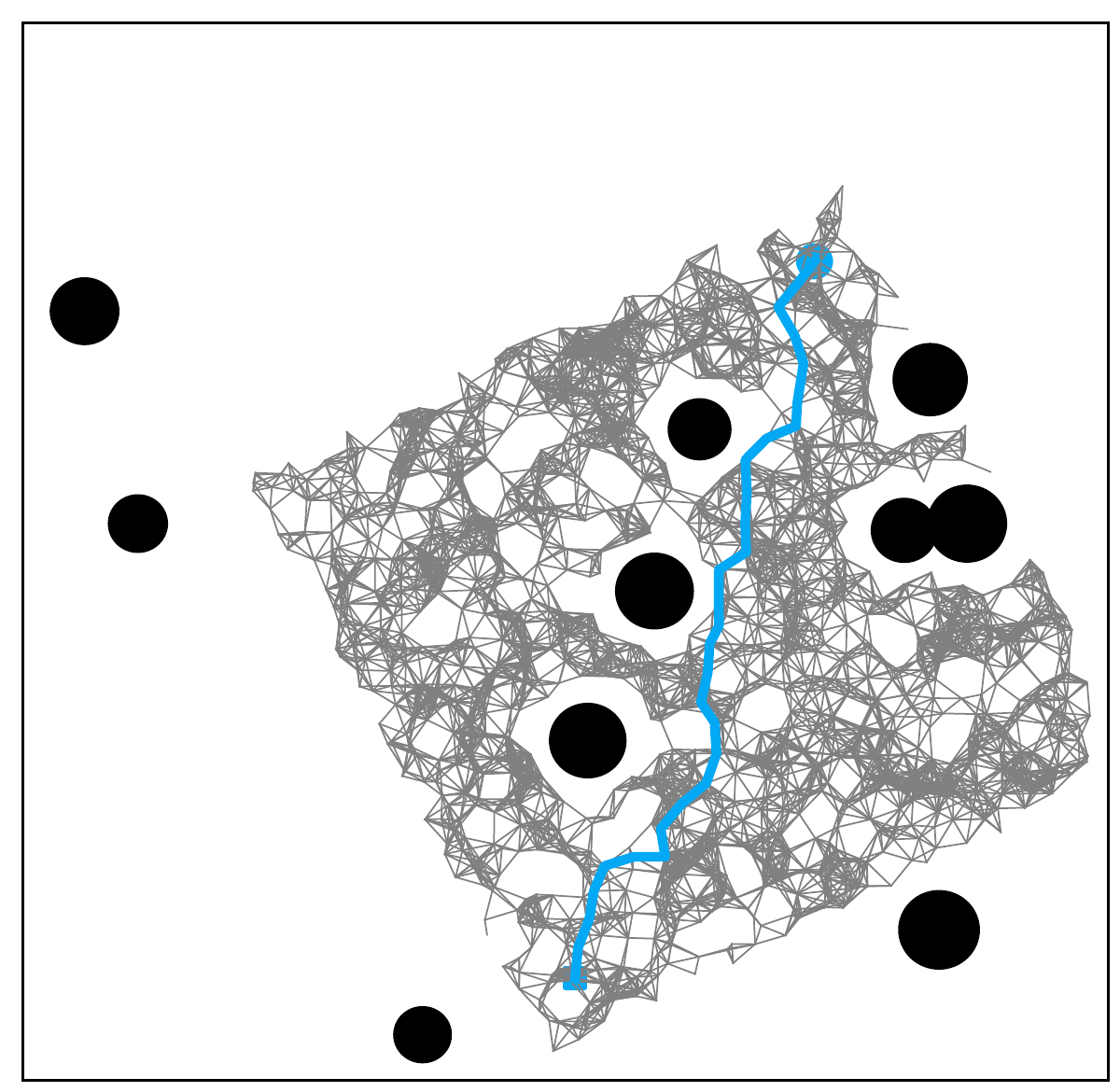}{square}
\end{tabular}
}
\caption{
Constructed roadmaps for one agent in the Basic scenario.
{\normalfont
The blue line is a path in a solution.
Parameters are: 
\textit{CTRMs}: $\ntraj=25$;
\textit{random}: 3000 samples;
\textit{grid}: $32\times 32$;
\textit{square}: the middle density.
}
}
\label{fig:roadmaps}
\end{figure*}
}

Once a vertex $(p, t)$ is sampled either from the model or randomly, we validate whether this is reachable from the current location $\locbody{i}{t-1}$ using the local planner. If $p$ is invalid, we repeat the random-walk sampling up to the maximum times specified for resampling, and return $p=\locbody{i}{t-1}$ (which is trivially valid) if $p$ is still invalid.

\subsection{Finding Compatible Vertices}
\label{sec:merge}
At Line~\ref{algo:traj:merge} of Alg.~\ref{algo:roadmap-generation}, we invoke \texttt{find\_compatible\_vertex} to search the current CTRMs for a vertex that is compatible with the sampled vertex $(p, t)$. By replacing the original vertex $(p, t)$ with a compatible one $(q, t)\; \in V_i$, we can reduce the time required for connectivity checking in the \funcname{insert} function and, more importantly, reduce the search space for multi-agent path planners. The specific algorithm is presented in Sec.~B.2 of the \supp. Intuitively, two vertices $p, q$ are regarded as compatible when they (1)~are spatially close enough and (2)~share the same connectivity to other vertices.
\section{Evaluation}
\label{sec:evaluation}
In this section, we evaluate CTRMs in various MAPP problems and clarify the merits of our method.

\subsection{Experimental Setups}

\subsubsection{MAPP Problems}
Recall that an MAPP problem instance is a tuple $\mathcal{I}=(\mathcal{A}, \world, \mathscr{O}, \mathscr{R}, \mathscr{K}, \mathcal{S}, \mathcal{G})$ (see Sec.~\ref{sec:pre}). For all the MAPP problem instances, we model agent bodies $\mathscr{R}$ and obstacles $\mathscr{O}$ as a circle in the 2D closed continuous space ($\world$ = $[0, 1]^2$) to detect collisions easily; hence, an agent body $\agentregion_i \in \mathscr{R}$ is characterized by a scalar value $r_i$ representing the radius. As kinematic constraints $\motion_i\in\mathscr{K}$, we assume that each agent has the maximum velocity $k_i$ and moves linearly in two vertices (space-time pairs) at constant velocity. With these settings, we generated 100 different instances, hereafter referred to as \emph{test} instances, for each of the following five scenarios of MAPP (see also the top of Fig.~\ref{fig:result}).
{
\smallskip
\newcommand{\myitemize}[2]{\noindent$\blacktriangleright$~{\emph{#1:}}~#2\smallskip}

\myitemize{Basic scenario}{A baseline scenario referred to as \textbf{(1) Basic} that corresponds to a discrete setting of a $32\times 32$ grid and contains 21--30 homogeneous agents as $\mathcal{A}$ and ten non-uniform obstacles as $\mathscr{O}$. For each timestep, an agent can move a maximum distance of $k_i=1/32$. The radius of agents is set to half of their maximum speed, \ie, $r_i=1/64$. For each problem instance, the number of agents $N$ was randomly determined from the range $N\in \{21,22,\ldots, 30\}$. The initial positions $\mathcal{S}$, goal positions $\mathcal{G}$, and the positions of the obstacles were set randomly for each instance}

\myitemize{Variants of the basic scenario}{Three scenarios that each change one of the settings from those of \textbf{Basic}: \textbf{(2) More Agents} with $N\in\{31,32,\ldots,40\}$, \textbf{(3) No Obstacles} with no obstacles, and \textbf{(4) More Obstacles} with 20 non-uniform obstacles as $\mathscr{O}$, to see how robust the proposed approach against these parameters.}

\myitemize{Heterogeneous agent scenario}{An advanced scenario called \textbf{(5) Hetero Agents} that contains $N\in\{21,22,\ldots,30\}$ agents whose size $r_i$ and speed $k_i$ are each multiplied randomly by $\times 1$, $\times 1.25$, or $\times 1.5$ from the original ones defined in \textbf{Basic}.}
}

\subsubsection{Baselines for Roadmap Construction Methods}
To evaluate the effectiveness of CTRMs, we implemented four other roadmap construction methods (see also Fig.~\ref{fig:roadmaps}) that are non-timed and non-cooperative, which have commonly been used for single-agent SBMP and previous work on MAPP in continuous spaces.

{
\smallskip
\newcommand{\myitemize}[1]{\noindent$\blacktriangleright$~#1\smallskip}

\myitemize{\textbf{Random sampling (random)} that samples agent locations uniformly at random from the space. This is equivalent to a simplified version of PRM~\cite{karaman2011sampling} and has been used as part of the procedure to solve MAPP~\cite{van2005prioritized,solis2021representation}. The numbers of samples were set to $\{3000, 5000, 7000\}$.}
 
\myitemize{\textbf{Grid sampling (grid)} like the ones the conventional MAPF assumes as a discretized environment~\cite{stern2019def}. The grid sizes were set to $\{32\times 32, 64\times 64, 84\times 84 \}$.}

\myitemize{\textbf{SPARS (SPArse Roadmap Spanner algorithm)}~\cite{dobson2014sparse}, an algorithm for roadmap construction that attempts to reduce both vertices and edges. SPARS has been developed for single-agent planning and has also been used in MAPP~\cite{honig2018trajectory}. We used the implementation in the Open Motion Planning Library~\cite{sucan2012ompl}.}

\myitemize{\textbf{Square sampling (square)} as a variant of the random sampling focusing on the square region with its diagonal line given by the start and goal for a single agent (with a margin). We introduce this sampling as a simpler approach to provide a set of agent-specific roadmaps. The number of samples was determined by the length of the diagonal line times a parameter given adaptively to generate low-, middle-, and high-density roadmaps.}
}

More details for SPARS and square sampling are presented in Sec.~C.1 of the \supp. Given a set of samples as vertices, we built roadmaps by creating edges at two vertices where agents can travel within a single timestep given their size $r_i$ and speed $k_i$ parameters. For the former three methods, a single roadmap was shared across all the agents in the scenarios (1--4) since their size and speed were set identically. For scenario (5) with heterogeneous agents, we constructed different roadmaps for individual agents, as done for CTRMs and the square sampling.  

{
\setlength{\tabcolsep}{0pt}
\newcommand{\spacing}{\vspace{0.23cm}}
\newcommand{\figcol}[5]{
\begin{minipage}[t]{0.185\linewidth}
\centering
{\small\textbf{#5}}\\
{\footnotesize #2, #3, #4}\\
\includegraphics[width=0.8\linewidth,right]{fig/raw/instance/instance_solution_#1.pdf}\\
\spacing
\includegraphics[width=1\linewidth,right]{fig/raw/success-rate/success_rate_vs_sample_nums_per_agent_per_timestep_#1.pdf}\\
\spacing
\includegraphics[width=1\linewidth,right]{fig/raw/solution-quality/cost_vs_expanded-nodes_#1.pdf}\\
\spacing
\includegraphics[width=1\linewidth,right]{fig/raw/runtime/runtime_stack_#1.pdf}
\end{minipage}
}

\begin{figure*}
\centering
\begin{tabular}{ccccc}
\figcol{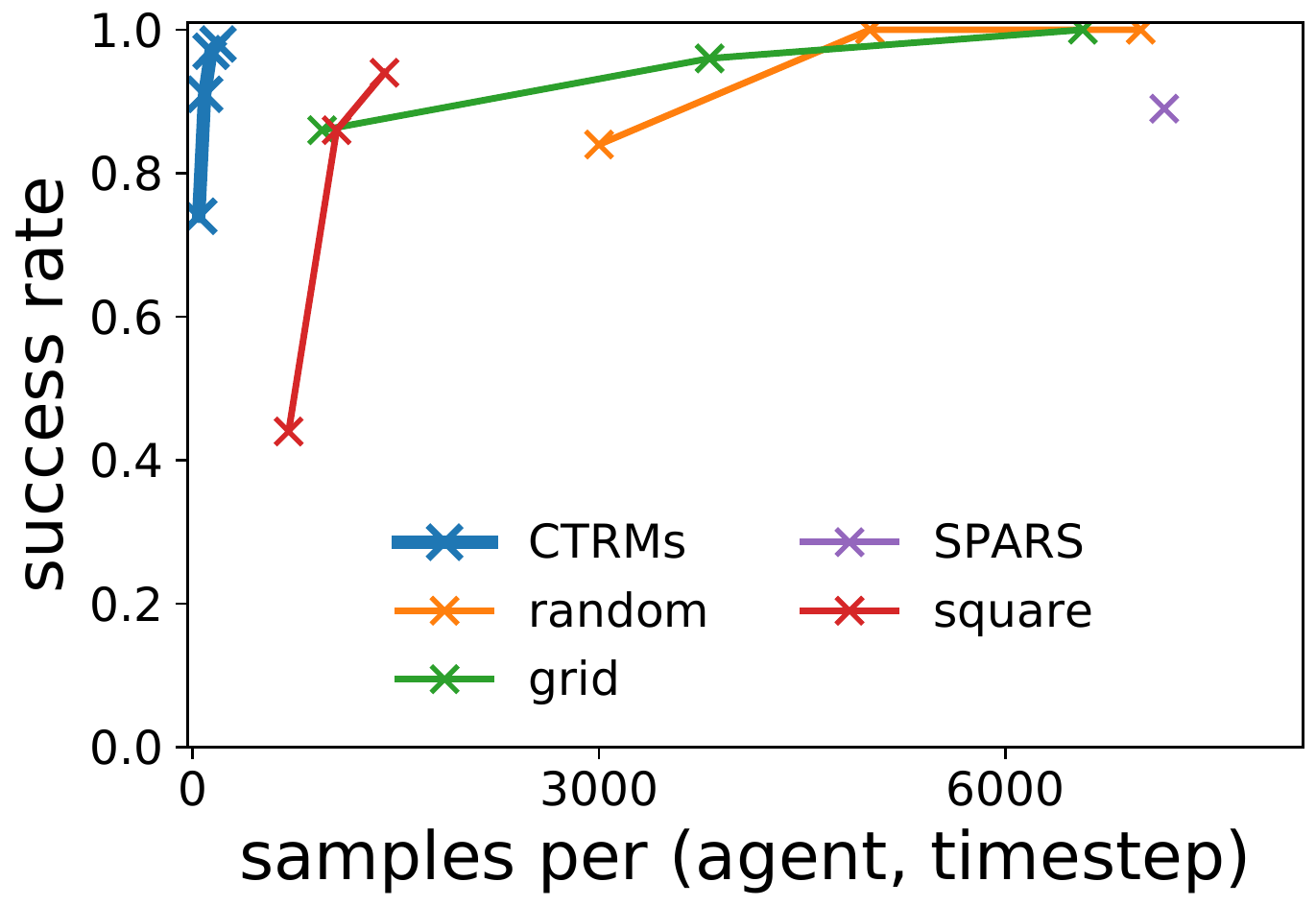}{homo}{21-30 agents}{10 obs}{(1)~Basic}
\figcol{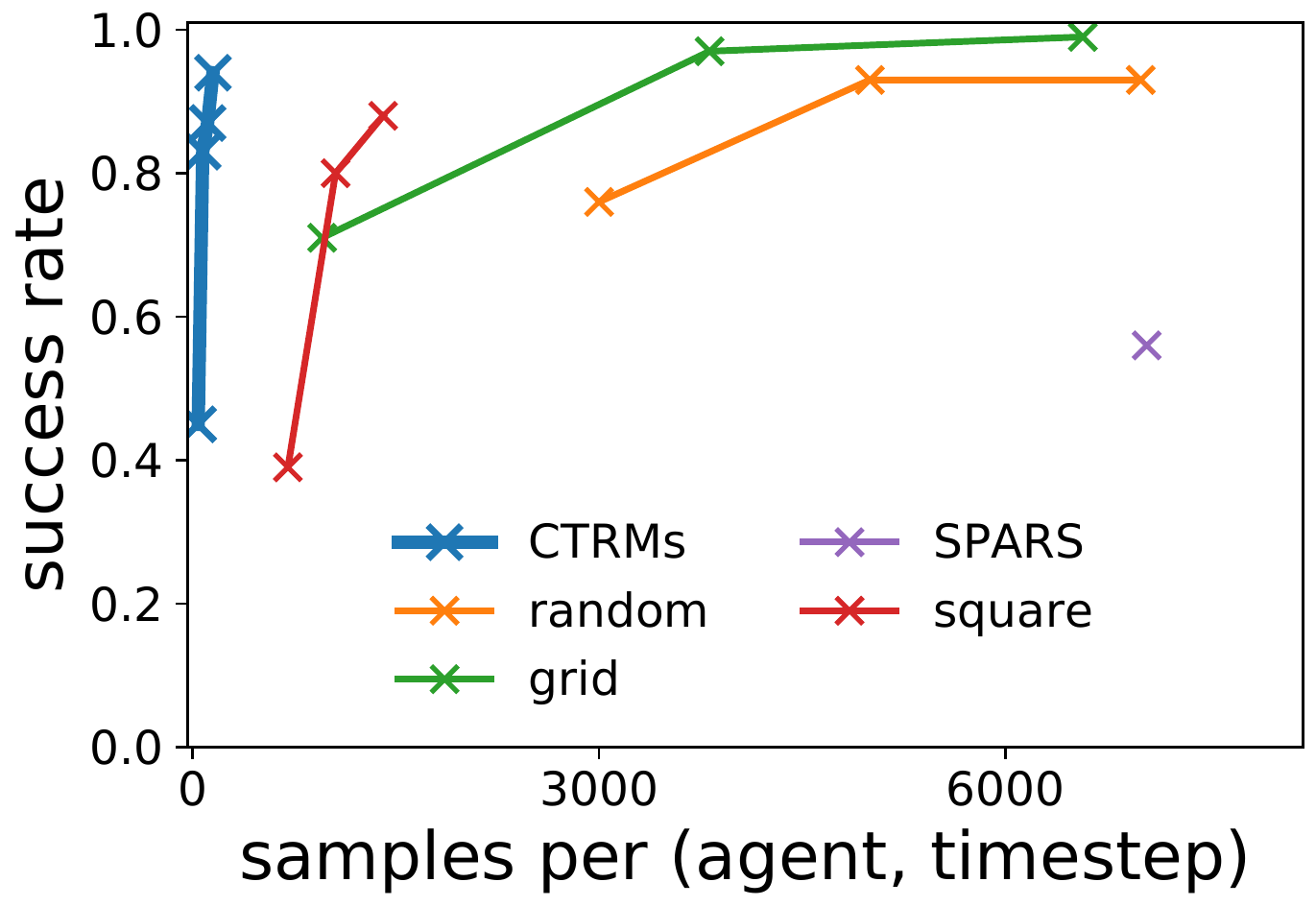}{homo}{\textbf{31-40 agents}}{10 obs}{(2)~More Agents}
\figcol{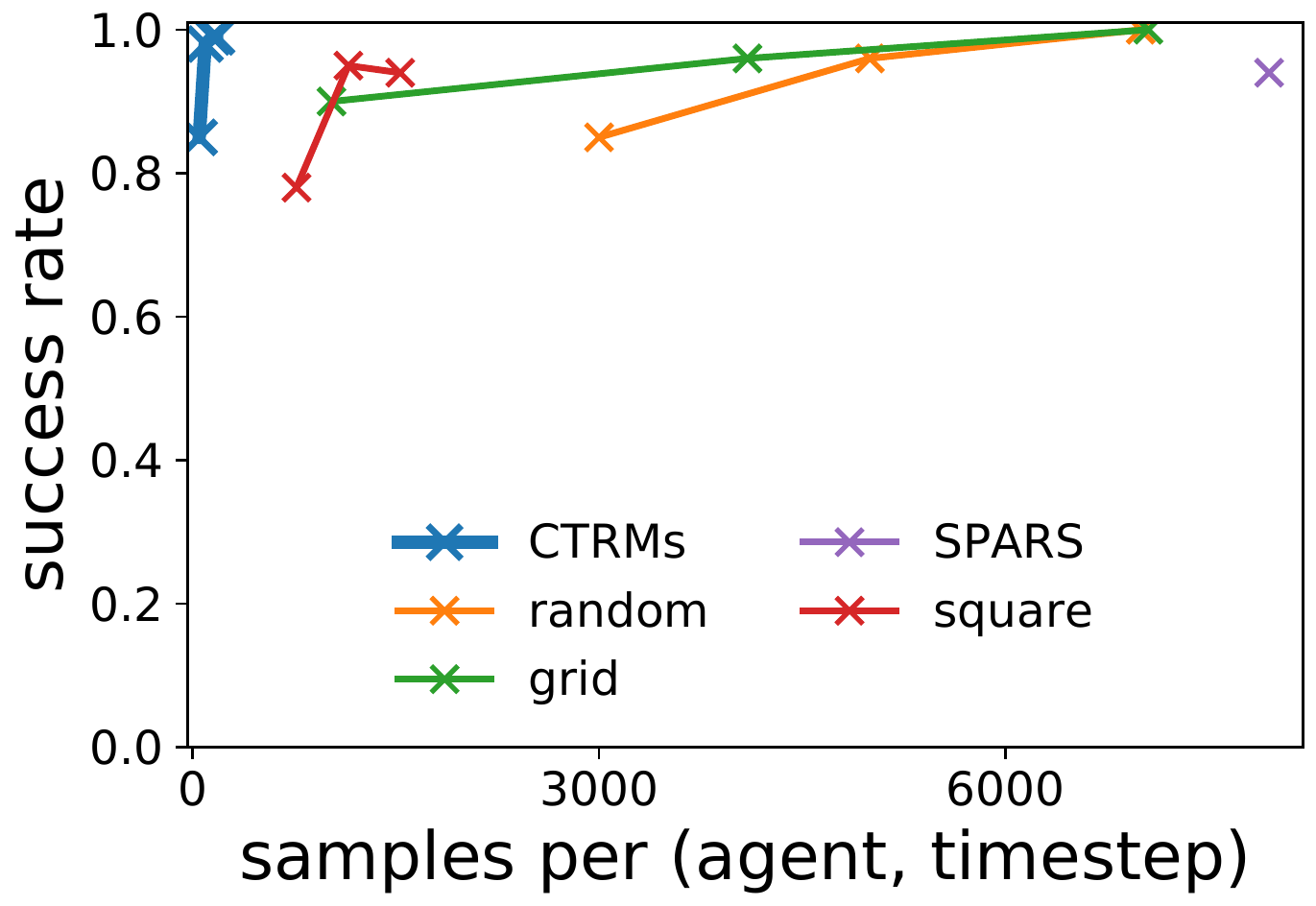}{homo}{21-30 agents}{\textbf{0 obs}}{(3)~No Obstacles}
\figcol{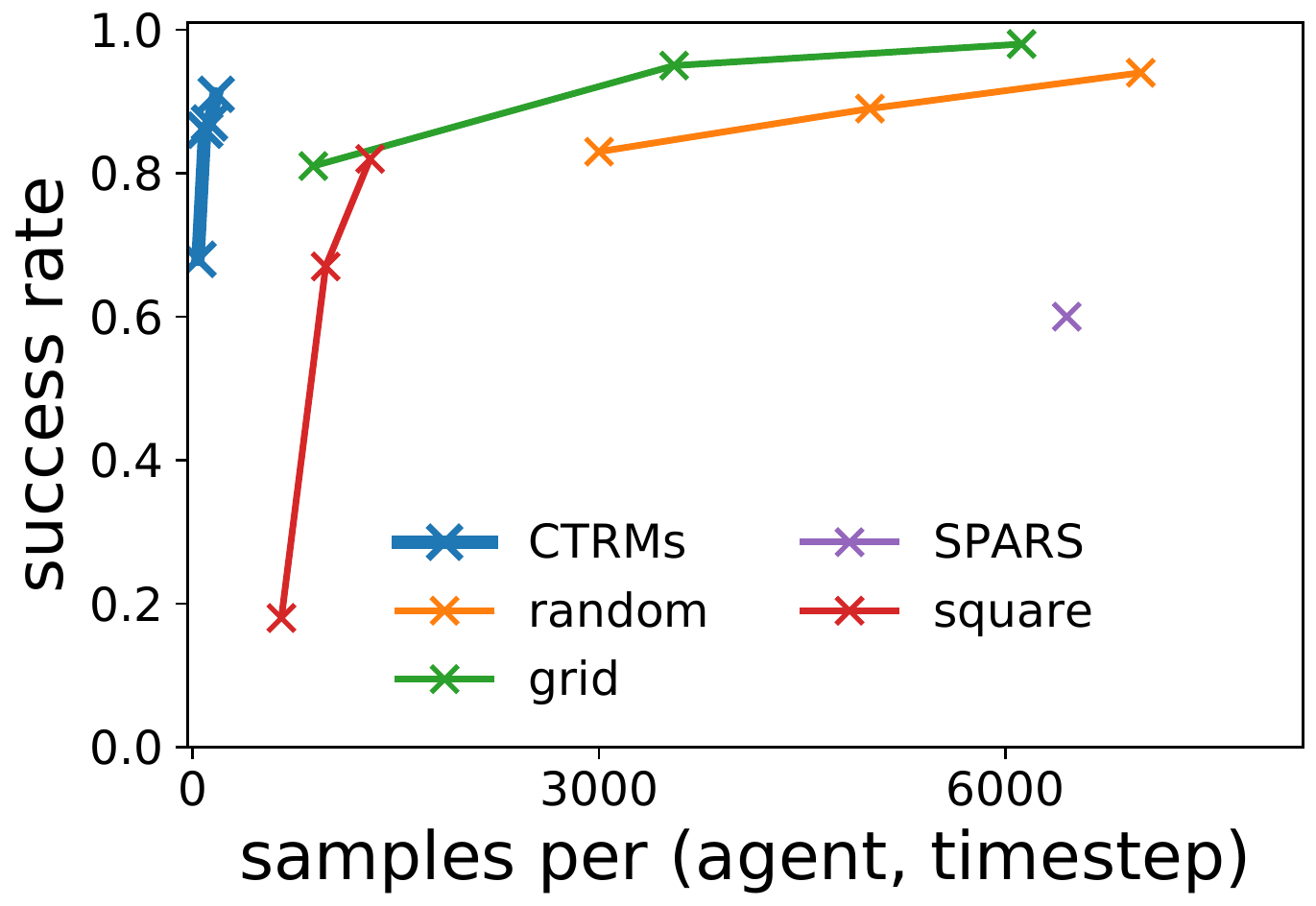}{homo}{21-30 agents}{\textbf{20 obs}}{(4)~More Obstacles}
\figcol{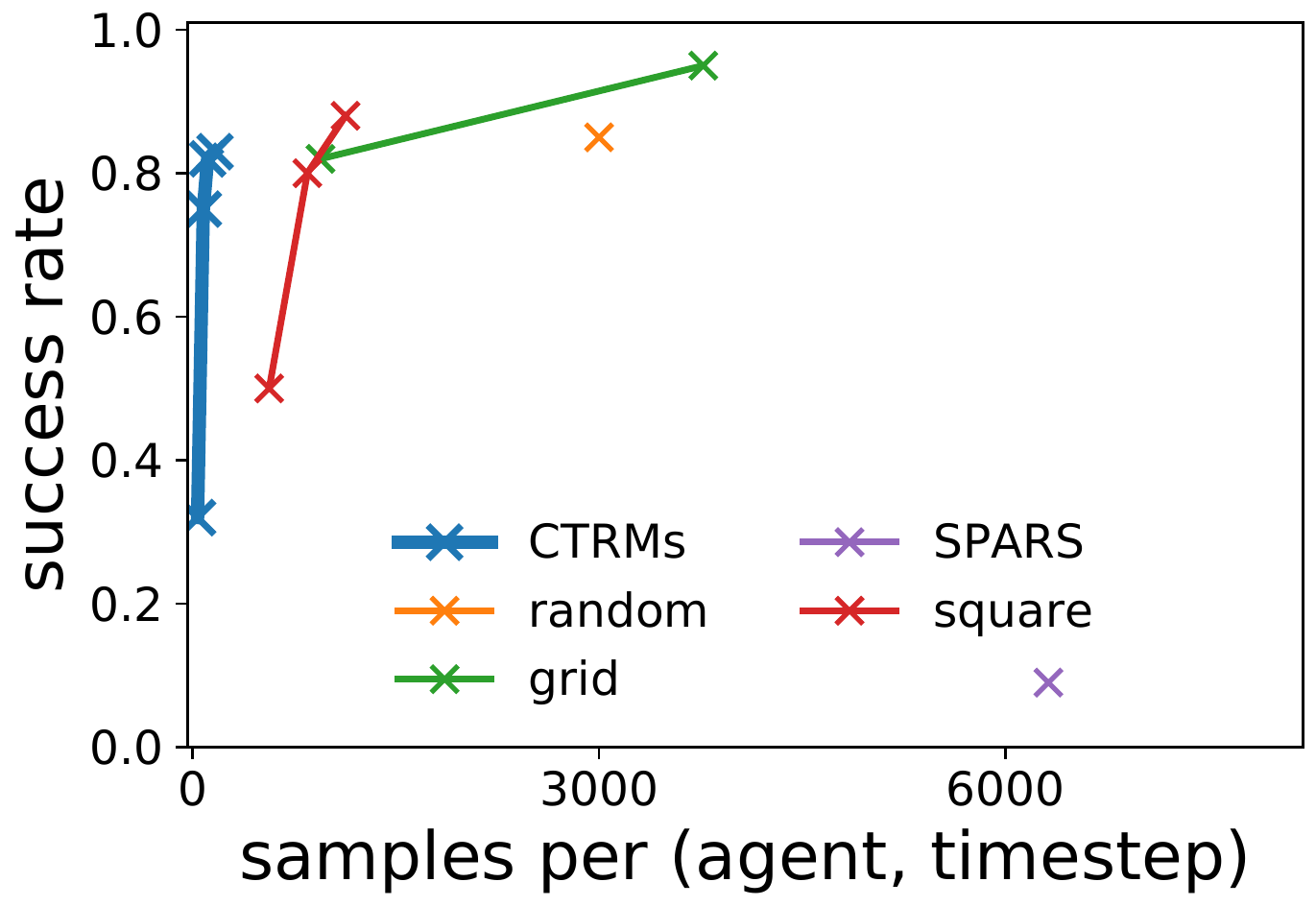}{\textbf{hetero}}{21-30 agents}{10 obs}{(5)~Hetero Agents}
\end{tabular}
\caption{Summary of results. {\normalfont The columns correspond to respective scenarios. For each method, plots to the right correspond to the denser roadmaps (\eg, CTRMs: $\ntraj=25 \rightarrow \ntraj=100$, grid: $32\times 32 \rightarrow 84 \times 84$). \textit{First row}: Solution examples planned by PP on CTRMs ($\ntraj=25$). \textit{Second row}: Success rate vs. the number of samples per agent and per timestep in the constructed roadmaps. \textit{Third row}: Sum-of-costs vs. the number of expanded search nodes in the planning phase. Each value is normalized by dividing by the number of agents. \textit{Fourth row}: Overall runtime, consisting of the planning and the roadmap construction phases. For the results in the third and fourth rows, in order to obtain meaningful insights, we excluded methods with success rates below 70\% and calculated the average of the metrics for problem instances where all the remaining methods were successful in planning. We also omitted some baselines with specific parameter settings resulting in extremely long computation times (over 15 minutes per problem instance) from the evaluation on the Hetero Agent scenario. All these omitted methods and settings are marked with ``x'' in the fourth row.}
}
\label{fig:result}
\end{figure*}
}

\subsubsection{Evaluation Metrics}
Since the objective of MAPP is to plan conflict-free trajectories, the effectiveness of roadmap construction methods should be evaluated in terms of the subsequent planning results. We used standard prioritized planning (PP)~\cite{silver2005cooperative,van2005prioritized} as the planner and computed the following metrics.
{
\smallskip
\newcommand{\myitemize}[2]{\noindent$\blacktriangleright$~\textbf{#1}:~#2\smallskip}

\myitemize{Success rate of the planning}{Percentage of successful planning among 100 instances. We regard two cases as a failure: when the PP yielded failure or when the planning time reached a 10-minute timeout.}

\myitemize{Sum-of-costs}{Solution quality defined in Sec.~\ref{sec:pre} averaged across all problem instances that resulted in planning success.}

\myitemize{The number of expanded search nodes in the planning}{Metric of the planning effort (the smaller, the better), which was also averaged across all problem instances that resulted in planning success. We use this metric as a proxy of runtime, since actual runtimes rely heavily on implementations. Specifically, we simply counted the number of search nodes expanded in PP.}

\myitemize{Runtime}{Reference record of the execution time for the roadmap construction and the subsequent planning.}
}

Note that we implemented all the methods in Python with partial use of C++. The results were obtained on a desktop PC with Intel Core i7-8700 CPU and 32GB RAM.

\subsection{Implementation Details for CTRMs}
\label{sec:impl}

\paragraph{Model Architectures} All neural networks used in our method are standard multilayer perceptrons (see Sec.~C.2 in the \supp for details). The FOV was set to $l=19$ and $L=160$. The number of neighboring agents of $\neigh{i}$ was set to 15 based on the distance between the current locations of agents, $\loc{i}{t}$. The indicator feature $\vct{x}\sub{ind}$ was defined by a three-dimensional one-hot vector indicating left, straight, and right, which were determined based on whether the sine of the angle between two vectors $g_i - \loc{i}{t}$ and $\loc{i}{t+1} - \loc{i}{t}$ was in $[-1, -\frac{1}{3}]$, $(-\frac{1}{3}, \frac{1}{3}]$, or $(\frac{1}{3}, 1]$.

\paragraph{Training Setups} We created 1,000 problem instances for training data and 100 instances for validation, where the trained model with the minimum loss on the validation data was stored and used to construct CTRMs for test instances. Regardless of test scenarios, we used the same training and validation data so that the parameters followed the \textbf{Hetero-Agent} scenario. This is because the Hetero-Agent scenario naturally contains diverse agent behaviors that could also be observed in the other scenarios, which helps reduce the time needed for training with little empirical performance degradation. Note that no identical instances were shared among training, validation, and test instances. Solutions to those instances were obtained by means of roadmap construction based on random sampling with 3,000 samples and path planning by PP. Model parameters were updated using the Adam optimizer~\cite{kingma2014adam} with the mini-batch size, number of epochs, and learning rate set to $(50, 1000, 0.001)$. We conducted the data creation and model training on a single workstation equipped with an Intel Xeon CPU E5-2698 v4 and NVIDIA Tesla V100 GPU to complete the procedures in a reasonable amount of time (data creation: 1 day with $\times 40$ multiprocessing; training: 2 hours).

\paragraph{Weighted Loss} We observed that the training and validation data created as above contained many agents that trivially moved along the shortest path from their start to goal positions, while we expect to observe actions of agents dodging each other to learn their interactions. As such data imbalance typically makes training difficult~\cite{lin2017focal,lu2018deep}, we introduced a weighted loss technique that gives different weights to each training sample based on the angle $\Delta$ between $g_i - \loc{i}{t}$ and $\loc{i}{t+1} - \loc{i}{t}$ with the following criterion: $1 - \exp \left(- \gamma\Delta^2\right)$, where $\gamma=50$. Intuitively, we gave smaller weights to samples that just move agents forward. 

\paragraph{Roadmap Construction} We set the parameters for constructing CTRMs in Alg.~\ref{algo:roadmap-generation} as follows. \ntraj was set to $\{25, 50, 75, 100\}$, and \tmax was $64$. In \funcname{sample\_next\_vertex}, we set \pbiased adaptively to timestep $t$ with $1 - \exp\left(-\gamma t / \min(\tmax, \tmakespan)\right)$, where $\gamma=5$, until each agent arrives at its goal. \pbiased was then fixed to 0.1 after agents reach their goal so that they can move around to avoid other agents still moving toward their goal.

\subsection{Results}
Figure~\ref{fig:result} shows our main results, and examples of constructed roadmaps are provided in Fig.~\ref{fig:roadmaps}. Overall, we confirmed that our approach substantially reduced the planning effort to obtain plausible solutions while maintaining a high success rate and sum-of-costs. Specific findings are summarized below.
{
\smallskip
\newcommand{\myitemize}[2]{\noindent$\blacktriangleright$~{\emph{#1}}~#2\smallskip}

\myitemize{CTRMs contain solutions in a small search space.}{The size of the search space can be assessed by the number of vertices in each roadmap per agent and per timestep, as compared to the success rate in the second row of Fig.~\ref{fig:result}. CTRMs have significantly fewer vertices compared to the others, but still allow the subsequent planner to successfully find a solution at a high rate. The roadmaps created by the random, grid, and SPARS sampling provided much larger search spaces since they are shared among agents and should thus cover the entire space. The square sampling was a bit better than those baselines, but was consistently outperformed by CTRMs.}

\myitemize{While keeping solution quality, CTRMs contribute to reducing the planning effort by several orders of magnitude}{compared to the baselines, as indicated in the third and fourth rows, simply thanks to the small search space. The sum-of-cost metric normalized by the number of agents was comparable to the baseline methods and even better than when the number of samples for the baseline methods is limited. This is primarily because our approach learns where to sample vertices from the solution paths of MAPP demonstrations.}

\myitemize{CTRMs achieve efficient MAPP solving from the end-to-end perspective.}{The total runtime with the overhead from roadmap constructions is shown in the fourth row. Although the runtime relies heavily on the implementation, our method can produce CTRMs in a realistic computation time. Once CTRMs are constructed, the planning can be finished immediately, unlike the baselines, despite the use of the same planning algorithm and implementation. Even so, there is room for further optimization in the roadmap construction, including improvement to the implementation of connectivity checks between vertices, as it is often a bottleneck of SBMP~\cite{elbanhawi2014sampling}.}

\myitemize{Our approach is generalizable for various scenarios.}{Even though our model was trained on just a single scenario with heterogeneous agents, the model once trained can be used for other scenarios with different numbers of agents and obstacles.}
}

\paragraph{Ablation Study} We further assessed the effect of each technical component used in our approach. Specifically, we evaluated degraded versions of our approach that omit 1) communication features (no $\vct{x}\sub{comm}$) or 2) indicator features (no $\vct{x}\sub{ind}$) when learning the model, and 3) no random walk (\ie, \pbiased=1) when constructing CTRMs. All the variants were evaluated on the Basic scenario in terms of success rate, sum-of-costs, and the number of expanded nodes with $N\sub{traj}=25$. As shown in Table~\ref{table:ablation}, involving the random walk is particularly critical to ensure a high success rate. This is consistent with the common practice of previous studies~\cite{ichter2018learning,ichter2020learned,chen2020learning} that use learned sampling and random sampling together. On top of that, the lack of communication feature $\vct{x}\sub{comm}$ greatly limits the success rate, sum-of-costs, and the number of expanded nodes, as agents are then required to avoid others solely by random walk. Furthermore, we confirmed that the indicator feature $\vct{x}\sub{ind}$ is equally important to improve the overall performance.

\begin{table}
 \caption{ Results of ablation study performed on the Basic scenario. {\normalfont Sum-of-costs and expanded nodes are normalized by the number of agents and averaged over 11 instances that succeeded in all methods except ``no random walk'' (resulting in an extremely low success rate), accompanied by 95\% confidence intervals. }}
  \label{table:ablation}
  \vspace{-0.3cm}
  \centering
  {
    \setlength{\tabcolsep}{3pt}
    
    \scalebox{0.9}{
    \begin{tabular}{lccc}
      \toprule
      & Success rate & Sum-of-costs & Expanded nodes \\
      \midrule
      CTRMs {\small ($\ntraj=25$)} & \textbf{0.80} & \textbf{21.2 {\small (20.3, 22.0)}} & \textbf{612.4 {\small (547.4, 674.2)}}\\
      no $\vct{x}\sub{comm}$ & 0.23 & 28.7  {\small (27.3, 30.2)} & 996.3  {\small (923.1, 1068.5)}\\
      no $\vct{x}\sub{ind}$ & 0.33 & 31.3 {\small (30.5, 32.2)} &  1058.6 {\small (993.8, 1117.3)}\\
      no random walk & 0.03 & N/A & N/A \\
      \bottomrule
    \end{tabular}
  }
  }
\end{table}

\paragraph{Limitations} Although CTRMs generally worked well in various scenarios, we informally confirmed that the method sometimes failed in ``bug trap'' situations where SBMP generally struggles~\cite{lavalle2006planning}. In general, our approach is applicable only if each agent can find an obstacle-free path to reach the goal when constructing CTRMs. Failing to do so will make it impossible to perform MAPF in the later stage. One possible resolution would be to introduce bi-directional path generation like a variant of RRT~\cite{kuffner2000rrt}.  Improving the overall performances by replacing the multi-agent path planner from PP is a promising direction for future work.
\section{Conclusion}
\label{sec:conclusion}

This paper presented cooperative timed roadmaps (CTRMs), a novel graph representation tailored to multi-agent path planning in continuous spaces. The CTRMs are constructed to provide the subsequent planner with a small search space for each agent while simultaneously being aware of other agents to avoid potential inter-agent collisions and contain plausible solution paths. This can be done by learning a generative model from relevant demonstrations and using it as an effective vertex sampler. The experimental results demonstrate the effectiveness of CTRMs for a variety of MAPP problems. Future work will seek to extend the proposed method to anytime planning, planning in continuous time, or multi-agent motion planning in higher-dimensional spaces.

\section*{Acknowledgments}
We are grateful to Kazumi Kasaura for his insightful comments throughout this study.

\bibliographystyle{ACM-Reference-Format}
\bibliography{ref}

\section*{Appendix}
\medskip
\renewcommand{\thesection}{\Alph{section}}
\setcounter{section}{0}

\section{Conditional Variational Autoencoder and Its Training}
\label{sec:cvae}
\paragraph{Basic Formulation of CVAE} The objective of CVAE~\cite{sohn2015learning} is to approximate a conditional probability distribution $p(\vct{y}\mid\vct{x})$ of vector $\vct{y}$ given another vector $\vct{x}$. CVAE consists of \emph{encoder} ${\it Enc}(\vct{x};\theta)$ and \emph{decoder} ${\it Dec}(\vct{x},\vct{z};\phi)$, which are typically neural networks parameterized by $\theta$ and $\phi$, respectively. The encoder ${\it Enc}(\vct{x};\theta)$ takes $\vct{x}$ as input to produce a conditional probability distribution $p_\theta(\vct{z}\mid \vct{x})$, where $\vct{z}$ is a latent variable that represents $\vct{x}$ in a low-dimensional space called latent space. Here, we model $p_\theta$ by a discrete, categorical distribution following \cite{ivanovic2021multimodal}, but it is also possible to consider a continuous distribution (such as Gaussian distribution). As for the decoder ${\it Dec}(\vct{x},\vct{z};\phi)$, it receives a concatenation of $\vct{x}$ and $\vct{z}$ sampled from $p_\theta$ to approximate another conditional probability distribution $p_\phi(\vct{y}\mid \vct{x},\vct{z})$. By marginalizing out $\vct{z}$, we can obtain $p(\vct{y}\mid\vct{x})$ as follows.
\begin{align}
  p(\vct{y}\mid\vct{x}) = \sum_{\vct{z}} p_\phi(\vct{y}\mid\vct{x}, \vct{z}) p_\theta(\vct{z}\mid\vct{x})
  \label{eq:p_y_x}
\end{align}

{
  \begin{figure}
    \centering
    \includegraphics[width=0.8\linewidth]{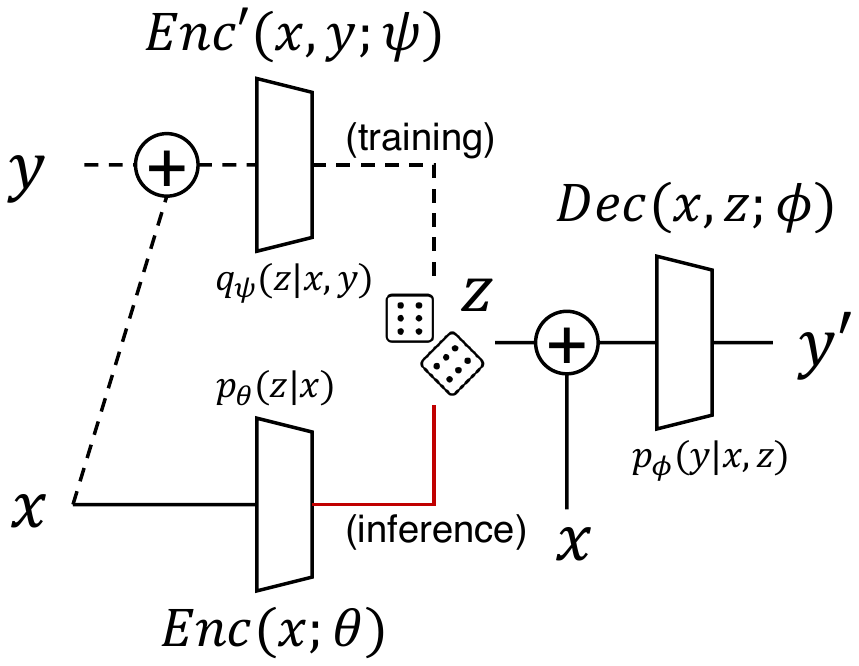}
    \caption{CAVE architecture.}
    \label{fig:cvae}
  \end{figure}
}

\paragraph{CVAE with Importance Sampling} To perform the marginalization of Eq.~(\ref{eq:p_y_x}), we expect $\vct{z}$ to contain some information about $\vct{y}$, otherwise $p_\theta(\vct{z}\mid\vct{x})$ will contribute very little to $p(\vct{y}\mid\vct{x})$. A typical approach to this problem is \emph{importance sampling}~\cite{bishop2006pattern}, where we sample $\vct{z}$ from a \emph{proposal distribution} $q(\vct{z}\mid\vct{x}, \vct{y})$ to select a proper $\vct{z}$. To this end, we introduce another neural network-based encoder ${\it Enc'}(\vct{x},\vct{y};\psi)$ parameterized by $\psi$, which takes the concatenation of $\vct{x}$ and $\vct{y}$ to produce the proposal distribution $q_\psi(\vct{z}\mid\vct{x},\vct{y})$. Equation~(\ref{eq:p_y_x}) can then be rewritten as follows:
\begin{equation}
  \begin{split}
 p(\vct{y}\mid\vct{x})
 &= \sum_{\vct{z}} \frac{p_\phi(\vct{y}\mid\vct{x},\vct{z}) p_\theta(\vct{z}\mid\vct{x})}{q_\psi(\vct{z}\mid\vct{x},\vct{y})} q_\psi(\vct{z}\mid\vct{x}, \vct{y}) \\
 &= \expectation{q_\psi(\vct{z}\mid\vct{x}, \vct{y})}
 {\frac{p_\phi(\vct{y}\mid\vct{x}, \vct{z}) p_\theta(\vct{z}\mid\vct{x})}{q_\psi(\vct{z}\mid\vct{x}, \vct{y})}}
  \end{split}
  \label{eq:expectation}
\end{equation}
By taking the $\log$ of both sides and using Jensen's inequality, we can derive the \emph{evidence lower-bound (ELBO)} on the right-hand side in the following inequality:
\begin{equation}
  \begin{split}
 \log p(\vct{y}\mid\vct{x}) \geq~
 &\expectation{q_\psi(\vct{z}\mid\vct{x}, \vct{y})}{\log p_\phi(\vct{y}\mid\vct{x}, \vct{z})}\\
 &- \KL{q_\psi(\vct{z}\mid\vct{x}, \vct{y})}{p_\theta(\vct{z}\mid\vct{x})}
  \end{split}
  \label{eq:elbo-trajectron}
\end{equation}
where $\KL{p}{q} = \expectation{p}{\log (p(\vct{x}) / q(\vct{x}))}$ is a Kullback-Leibler (KL) divergence.

\paragraph{Training CVAE} Given a collection of pairs of $\vct{x}$ and $\vct{y}$, we maximize this ELBO with respect to $\theta,\phi,\psi$ to achieve $p(\vct{y}\mid \vct{x})$. More specifically, the training of CVAE proceeds as follows (see also Fig.~\ref{fig:cvae}). First, the encoder ${\it Enc}(\vct{x};\theta)$ takes $\vct{x}$ as input to output the log of $p_\theta(\vct{z}\mid \vct{x})$. At the same time, the other encoder ${\it Enc'}(\vct{x},\vct{y};\psi)$ receives the concatenation of $\vct{x}$ and $\vct{y}$ to output the log of $q_\psi(\vct{z}\mid \vct{x},\vct{y})$. Then, a latent variable $\vct{z}$ is drawn from $q_\psi$, concatenated with $\vct{x}$, and fed to the decoder ${\it Dec}(\vct{x},\vct{z};\phi)$ to obtain $\vct{y}'$. This $\vct{y}'$ can be regarded as a sample drawn from $p(\vct{y}\mid\vct{x})$ under the current parameters $\theta,\phi,\psi$. Therefore, minimizing the discrepancy between $\vct{y}'$ and $\vct{y}$ corresponds to maximizing the first log-likelihood term of Eq.~(\ref{eq:elbo-trajectron}). Here, we leverage a reparameterization trick~\cite{kingma2014auto,eric2017categorical,chris2017concrete} for the sampling of latent variable $\vct{z}$ to enable the end-to-end learning of encoder ${\it Enc'}(\vct{x},\vct{y};\psi)$ and decoder ${\it Dec}(\vct{x},\vct{z};\phi)$ by means of back-propagation. Furthermore, it is easy to compute the second KL term of Eq.~(\ref{eq:elbo-trajectron}) directly from the outputs of ${\it Enc}(\vct{x};\psi)$ and ${\it Enc'}(\vct{x},\vct{y};\psi)$. In practice, we measure the L2 distance between $\vct{y}'$ and $\vct{y}$ for the first log-likelihood term, and give a scalar weight $0.1$ to the second KL term~\cite{ichter2018learning}, as it performed empirically well.

\paragraph{Using CVAE as a Sampler}
Once trained, CVAE in the inference time can be used as a conditional sampler taking $\vct{x}$ as input to generate plausible $\vct{y}'$, which is denoted by $F\sub{CTRM}(\vct{x})$ in our work. Note that we don't use ${\it Enc'}(\vct{x},\vct{y})$ any more for this sampling.

\paragraph{Joint Training with Other Network Modules}
As described in Sec.~\ref{sec:together}, we train the CVAE with the other network modules $\nn{self\_env}$, $\nn{other\_env}$, $\nn{comm}$, and $\nn{ind}$ jointly. Specifically, we calculate the negative log likelihood (NLL) loss between the softmax output of $\nn{ind}$ and the ground-truth values of $\vct{x}\sub{ind}$ computed exactly using solution paths in the training data. Then we minimize the multi-task loss defined by the sum of CVAE loss and NLL loss scaled by a factor of 0.001. This scaling was necessarly to roughly match the magnitude of the two losses. Since the outputs from $\nn{self\_env}$, $\nn{other\_env}$, and $\nn{comm}$ are used directly as inputs to the CVAE and $\nn{ind}$, the parameters of those modules can also be updated end-to-end by back-propagating the multi-task loss.

\section{Further Details for Constructing CTRMs}
\subsection{Sample Next Vertices}
\label{sec:sampling}
\begin{algorithm}[t]
  \caption{\funcname{sample\_next\_vertex}}
  \label{algo:get-sample}
  \begin{algorithmic}[1]
  \item[\textbf{Input}: instance \I, timestep $t$, agent $i$, location table \lochead]
  \item[\textbf{Output}: one sample $p \in \freespace{i}$]
  \item[\textbf{Hyperparameters}: $0 \leq \pbiased \leq 1$, $N\sub{retry} \in \mathbb{N}$]
    \If{with probability \pbiased}
    \State $p \sim F\sub{CTRM}$
    \label{algo:sampling:biased}
    \IfSingle{\funcname{valid\_edge}(\locbody{i}{t-1}, p, \I, i)}{\Return p}
    \EndIf
    \For{$k = 1...N\sub{retry}$}
    \State $p \leftarrow$~random walk from \locbody{i}{t-1}
    \label{algo:sampling:random}
    \IfSingle{\funcname{valid\_edge}(\locbody{i}{t-1}, p, \I, i)}{\Return p}
    \EndFor
    \State \Return $\locbody{i}{t-1}$
    \label{algo:sampling:stay}
  \end{algorithmic}
\end{algorithm}

Algorithm~\ref{algo:get-sample} presents the complete procedure of \funcname{sample\_next\_vertex}. A sample is mainly obtained from the trained model $F\sub{CTRM}$ (Line~\ref{algo:sampling:biased}), but we also introduce a random walk centered at the last location (Line~\ref{algo:sampling:random}) with the probability of $1-\pbiased$. The validity of each sample is confirmed by the function \funcname{valid\_edge}, which is a black-box function assumed to return true if and only if two locations can be traveled for the given agent in the problem instance. If it fails to obtain valid samples using $F\sub{CTRM}$, we then try the random-walk sampling up a predefined number of times $N\sub{retry}=3$. If $p$ is still invalid even after the above resampling, the function returns \locbody{i}{t-1} (Line~\ref{algo:sampling:stay}), as it is a trivially valid location.

\newcommand{\vparents}{\m{V_{\text{parents}}}}
\newcommand{\vchildren}{\m{V_{\text{children}}}}

\begin{algorithm}[t]
  \caption{\funcname{find\_compatible\_vertex}}
  \label{algo:merging}
  \begin{algorithmic}[1]
  \item[\textbf{Input}:~instance \I, timestep $t$, agent $i$, location $p$, roadmap $D$]
  \item[\textbf{Output}:~location $q \in \freespace{i}$ or ``NOT FOUND'']
  \item[\textbf{Hyperparameters}: $\delta \in \mathbb{R}$, heuristics $h: \world \mapsto \mathbb{R}$]
    \State $\vparents \leftarrow \funcname{get\_parents\_candidates}(p, \I, t, i, D)$
    \label{algo:merge:get-parents}
    \State $\vchildren \leftarrow \funcname{get\_children\_candidates}(p, \I, t, i, D)$
    \label{algo:merge:get-children}

    \medskip
    \For{$v = (q, t^\prime) \in V_i$ s.t. $t^\prime = t$}
    \label{algo:merge:check-start}
    \IfSingle{$\|q - p\| > \delta$}{\textbf{continue}}
    \label{algo:merge:distance}

    \medskip
    \If{$\vparents = \text{parents of}~v \land \vchildren = \text{children of}~v$}
    \label{algo:merge:same-start}
    \If{$h(p) < h(q)$}
    \State replace $q$ of $v$ by $p$
    \State \Return $p$
    \Else
    \State \Return $q$
    \EndIf
    \EndIf
    \label{algo:merge:same-end}

    \medskip
    \If{$\vparents \subseteq \text{parents of}~v \land \vchildren \subseteq \text{children of}~v$}
    \label{algo:merge:merged-to-q-start}
    \State \Return $q$
    \EndIf
    \label{algo:merge:merged-to-q-end}

    \medskip
    \If{$\text{parents of}~v \subseteq \vparents \land \text{children of}~v \subseteq \vchildren$}
    \label{algo:merge:merged-to-p-start}
    \State replace $q$ of $v$ by $p$
    \State replace parents of $v$ by \vparents
    \State replace children of $v$ by \vchildren
    \State \Return $p$
    \EndIf
    \label{algo:merge:merged-to-p-end}

    \EndFor
    \label{algo:merge:check-end}

    \medskip
    \State \Return ``NOT FOUND''
    \label{algo:merge:not-found}
  \end{algorithmic}
\end{algorithm}

\subsection{Finding Compatible Vertices}
\label{sec:merge}
Algorithm~\ref{algo:merging} shows the specific algorithm we used for the sub-routine \funcname{find\_compatible\_vertex}, which is partially inspired by the post-processing technique of roadmap sparsification for single-agent motion planning~\cite{salzman2014sparsification}. Intuitively, two vertices $p, q$ are regarded as compatible when they (1)~are spatially close enough and (2)~share the same connectivity to other vertices. The former is determined by whether $\|q - p\| \leq \delta$ holds (Line~\ref{algo:merge:distance}), where $\delta \in \mathbb{R}$ was set to be the tenth of the maximum speed of each agent in our experiments. The latter needs to check the relationship of the parents and children of these samples. Specifically, the algorithm initially obtains potential parents and children for a new sample $p$ at timestep $t$ by two functions\\\funcname{get\_parents\_candidates} and \funcname{get\_children\_candidates} (Lines~\ref{algo:merge:get-parents}--\ref{algo:merge:get-children}), using the local planner. Then, for each vertex $v = (q, t) \in V_i$ of the current roadmap, the algorithm checks whether $p$ and $q$ are compatible. If so, it returns $p$ or $q$ based on their structures of parents and children (Lines~\ref{algo:merge:check-start}--\ref{algo:merge:check-end}), and otherwise returns ``NOT FOUND'' (Line~\ref{algo:merge:not-found}). We consider three cases for $p$ and $q$:
\begin{enumerate}
\item $p$ and $q$ have the same parents and children (Lines~\ref{algo:merge:same-start}--\ref{algo:merge:same-end}): Select one of them based on the heuristics $h(p)$, which in our experiment were given by the distance from $p$ to the goal.
 \item $q$ contains $p$'s edge structure (Lines~\ref{algo:merge:merged-to-q-start}--\ref{algo:merge:merged-to-q-end}): The algorithm simply returns $q$, as it can account for the edge structure of $p$.
 \item $p$ contains $q$'s edge structure (Lines~\ref{algo:merge:merged-to-p-start}--\ref{algo:merge:merged-to-p-end}): After replacing $q$ of $v$ with $p$ and updating the edge structure, the algorithm returns $p$.
\end{enumerate}

\section{Further Details of Experimental Setups}
\subsection{Parameters for Baselines}
\label{sec:baseline-params}

\paragraph{SPARS} The implementation in the Open Motion Planning Library~\cite{sucan2012ompl} has four hyperparameters: \texttt{dense\_delta\_fraction}, \\\texttt{spars\_delta\_fraction}, \texttt{stretch\_factor}, and time limits. We set these to $0.1$, $0.01$, $1.3$, and 30 seconds, respectively.

\paragraph{Square Sampling} The number of samples of each agent is determined by the length of the diagonal line $l$ (between its start and goal) times a given parameter $c$. Specifically, we determined the number by $cl / k_i$, where $k_i$ is the maximum velocity of agent $a_i$. We set $c$ to $50$ (low), $75$ (middle), and $100$ (high). The margin of the square region was set to $k_i/5$.

\subsection{Parameters for Our Method}

All the neural networks used in our method, \ie, \nn{self\_env}, \nn{other\_env}, \nn{comm}, \nn{ind}, along with the encoders ${\it Enc}(\vct{x};\theta)$, ${\it Enc}(\vct{x},\vct{y};\psi)$ and decoder ${\it Dec}(\vct{z};\phi)$ in CVAE, were defined by a standard multilayer perceptron with two fully connected layers and 32 channels, followed by ReLU activation~\cite{nair2010rectified} except for the last layer. The encoders and decoder also had batch normalization~\cite{ioffe2015batch} between the fully connected layers and ReLU to stabilize the training. The output of \nn{ind} was activated by the softmax function to predict one-hot indicator feature $\vct{x}\sub{ind}$, where the argmax function was further applied in the inference phase. The dimensions of message vectors $\vct{m}_{j\rightarrow i}$, attention vectors $\vct{\alpha}_{j\rightarrow i}$, and the latent variables of CVAE $\vct{z}$ were respectively set to 32, 10, and 64.
\end{document}